\documentclass{article}

\usepackage{epsfig}

\newcommand{\be}{\begin{equation}}
\newcommand{\ee}{\end{equation}}

\def \uu {{\bf u}}
\def \ww {{\bf w}}
\def \ff {{\bf f}}

\def \FF {{\bf F}}

\def \nn {{\bf n}}
\def \xx {{\bf x}}
\def \vv {{\bf v}}
\def \II {{\bf I}}

\def \JJ {{\bf J}}
\def \EE {{\bf E}}
\def \AA {{\bf A}}
\def \BB {{\bf B}}
\def \bb {{\bf b}}
\def \TT {{\bf T}}
\def \PPhi {{\bf \Phi}}

\def \su {\tilde u}
\def \sw {\tilde w}
\def \sB {\tilde B}
\def \sb {\tilde b}
\def \sD {\tilde D}
\def \Dx {\tilde {D}_x}
\def \Dy {\tilde {D}_y}
\def \Dz {\tilde {D}_z}
\def \sww {\tilde{\ww}}

\def \hb {{\hat b}}

\def \hA {{\hat A}}
\def \hE {{\hat E}}

\def \hff {{\hat \ff}}
\def \hv {{\hat v}}
\def \hJ {{J}}
\def \hJJ {{\JJ}}

\def \auu {{\overline\uu}}
\def \aFF {\overline{\FF}}
\def \aEE {\overline{\EE}}
\def \aww {\overline{\ww}}
\def \aff {\overline{\ff}}

\def \aE {\overline{E}}
\def \aA {\overline{A}}
\def \au {\overline{u}}
\def \ab {\overline{b}}
\def \aB {\overline{B}}

\def \au {\overline{u}}

\def \aB {\overline{B}}
\def \ab {\overline{b}}

\def \ad {\overline\delta}

\def \divb {{\nabla\cdot\BB}}
\def \divbb {{\nabla\cdot\bb}}
\def\divn {[\divb]_{num}}
\def\divnn {[\divbb]_{num}}

\def\xm {x_{j-1/2}}
\def\xp {x_{j+1/2}}
\def\ym {y_{k-1/2}}
\def\yp {y_{k+1/2}}
\def\zp {z_{m+1/2}}
\def\zm {z_{m-1/2}}
\def\xpm {x_{j\pm 1/2}}
\def\ypm {y_{k\pm 1/2}}

\def\jm {{j-1/2}}
\def\jp {{j+1/2}}

\def\kp {{k+1/2}}

\def\mp {{m+1/2}}
\def\jpm {{j\pm 1/2}}

\def\P {_{j,k,m}}

\def \fl {f^{(l)}}

\def\abx {\ab_x}
\def\aby {\ab_y}
\def\abz {\ab_z}

\def\fs  {f^*}
\def\Fs  {F^*}

\def\ffs {\ff^*}
\def\FFs {\FF^*}
\def\Es  {E^*}

\begin{document}

\title{On the Divergence-Free Condition in Godunov-Type \\
Schemes for Ideal Magnetohydrodynamics: \\
the Upwind Constrained Transport Method}

\author{P. Londrillo
\\INAF - Osservatorio di Bologna, londrillo@bo.astro.it
\and
L. Del Zanna
\\Universit\`a di Firenze, ldz@arcetri.astro.it}

\date{Accepted for publication in Journal of Computational Physics, Oct. 2003}

\maketitle

\begin{abstract}
We present a general framework to design Godunov-type schemes for
multidimensional ideal magnetohydrodynamic (MHD) systems, having the
divergence-free relation and the related properties of the magnetic
field $\BB$ as built-in conditions.
Our approach mostly relies on the {\em Constrained Transport} (CT)
discretization technique for the magnetic field components,
originally developed for the linear induction equation, which assures
$\divn=0$ and its preservation in time to within machine accuracy
in a finite-volume setting.
We show that the CT formalism, when fully exploited, can be used as a general
guideline to design the reconstruction procedures of the $\BB$ vector field,
to adapt standard upwind procedures for the
momentum and energy equations, avoiding the onset of numerical monopoles
of $O(1)$ size, and to formulate approximate Riemann solvers for the
induction equation. This general framework will be named here
{\em Upwind Constrained Transport} (UCT).
To demonstrate the versatility of our method, we apply it
to a variety of schemes, which are finally validated numerically
and compared: a novel implementation for the MHD case of the second order
Roe-type positive scheme by Liu and Lax (J. Comp. Fluid Dynam. 5,
133, 1996), and both the second and third order versions of a
central-type MHD scheme presented by Londrillo and Del Zanna
(Astrophys. J. 530, 508, 2000), where the basic UCT strategies
have been first outlined.
\end{abstract}

\section{Introduction}

In extending Godunov-type conservative schemes designed for Euler equations
of gas-dynamics to the system of (ideal) magnetohydrodynamics (MHD),
in the multidimensional case a main problem arises on how to represent
the solenoidal structure of the magnetic field vector $\BB$ and on how to
formulate reconstruction procedures and (approximate) Riemann
solvers sharing consistency with this property.
In the last years a number of works have focused on this specific problem
and many different approaches have been proposed.
A wide class of (second order) numerical schemes for regular grids have
been analyzed and compared by Toth \cite{T}, while contributions covering
also higher order schemes, adaptive mesh refinements (AMR) and unstructured
grids are in rapid development.

Since we are mainly interested here to analyze {\em methodological} aspects,
we propose a broad classification of the published contributions 
on this specific topic into {\em two main groups}:

\begin{enumerate}

\item
Schemes based on {\em standard upwind procedures} (henceforth SUP)
designed for Euler equations, where also magnetic field components are
discretized at cell centers as the other fluid variables. Since in this
case the approximated $\divn$ based on central derivatives may have a
non-vanishing size, different strategies to control or prevent the 
accumulation in time of related spurious numerical effects (usually 
referred to as {\em numerical monopoles}) have been proposed.

\begin{itemize}
\item
A first method, suggested by \cite{BB}, is to add an elliptic (Poisson)
equation to recover the solenoidal property at each time-step.
In reference \cite{T} this procedure has been named {\em projection scheme}
and is currently widely adopted (see \cite{JW} for a high order WENO scheme).
\item
In the scheme introduced by Powell \cite{P} (see also \cite{Petal}),
the numerical $\divn$ quantity is not forced to vanish; the MHD system
is reformulated by adding new source terms proportional to this variable
in order to recover the original MHD system in non-conservative form.
Moreover, the classical seven-mode Riemann wave fans have been enlarged
to eight modes. In this modified system, upwinding is applied to
all magnetic field components and hence also to the component
$B_n$ across a discontinuity surface.
\item
In a more recent work \cite{DKK}, in order to preserve both the conservative
form and the hyperbolic structure of the MHD system, a new time dependent
wave equation is introduced to damp and/or to transport away the
non-zero $\divn$ contributions.
\end{itemize}

\item
In the second group we include schemes which take advantage of the so-called
{\em Constrained Transport} (CT) method by Evans and Hawley \cite{EH}
(originally suggested for the evolution of the induction equation
in the linear approximation).
It is a main feature of this method to introduce staggered discretizations
of magnetic and electric vector fields in the induction equation.
In fact, by using these staggered values
to approximate the relevant first derivatives,
$\divn=0$ in the initial conditions and its exact preservation in time
result. The problem here is on how to apply this formalism in a
Godunov-type scheme for the full MHD system.

\begin{itemize}
\item
Most of the published works combine the above
CT discretization with the SUP cell centered discretization by
introducing different
empirical recipes (e.g. \cite{DW}, \cite{RMJF}, \cite{BS}, \cite{K}).
However, these procedures result in a sort of hybrid schemes and the problem
of numerical monopoles is still left open, in our opinion.

\item
In our previous work \cite{LD} (LD from now on) we have proposed
numerical procedures to take advantage of the specific CT
discretization benefits, and hence the $\divn=0$ condition,
even in the reconstruction steps and in the approximate Riemann solvers.
The same method has been then applied to relativistic MHD \cite{DBL}.
\end{itemize}

\end{enumerate}

The goal of the present paper is twofold. First, by adding
analytical arguments to the approach outlined in LD,
we propose {\em a method} to construct and then to characterize
{\em a class} of numerical schemes.
Second, we present implementations of a variety of different schemes,
to demonstrate the versatility and self-consistency of the method.

Regarding the first goal, our main concern is here to select a set of
properties, some of them common to the Euler system and other specific
of MHD equations, which in our opinion should be preserved in the
numerical discretization. In this way,
it is then possible to envisage Godunov-type schemes for MHD having:
(a) the divergence-free condition as an {\em exact built-in property},
(b) reconstruction and upwind procedures consistent with this property.
Since the CT formalism comes out to be the necessary starting
point to achieve this result, our framework will be named here
{\em Upwind Constrained Transport} (UCT) method.

As a novel numerical application we then propose the UCT implementation
of the {\em positive} scheme by Liu and Lax (\cite{LL1}, \cite{LL2}), a
second order Roe-type scheme which proves to be accurate and robust.
Numerical validation will be finally presented for
several standard two-dimensional test problems, where the results of
the new MHD positive scheme are compared with central-type schemes
as proposed in LD, extended here to more accurate central-upwind
two-speed approximate Riemann solvers, and tested in its second
and third order implementations.

This paper is organized as follows. In the next sub-section we propose
and discuss some general conditions as guidelines for numerical modeling.
The main ingredients to formulate general UCT-based Godunov-type schemes
for MHD systems, i.e. the discretization form, the proper reconstruction
procedures and the approximate Riemann solvers,
are presented in Sect.~2. In Sect.~3 we specify the method to the positive and
central MHD schemes, which will be finally tested and compared in Sect.~4.

\subsection{Conservation laws and consistency demands for numerical MHD}

The MHD system has a peculiar form and cannot be simply reduced to a set
of conservation laws for scalar variables, as the Euler equations.
In fact, if the specific structure of spatial differential operators
is taken into account, it is more properly represented by the set of
the following two coupled sub-systems:
\be
{\partial\uu \over \partial t}+\nabla\cdot\ff(\ww)=0,
\label{1a}
\ee
\be
{\partial \BB \over \partial t}+\nabla\times\EE(\ww)=0,
\label{1b}
\ee
equipped with the non-evolutionary constraint on the $\BB$ vector field
\be
\divb=0,
\label{1c}
\ee
which, once satisfied for initial conditions, is analytically preserved
in time by Eq.~(\ref{1b}).

The set of equations (\ref{1a}) evolves in time the five-component array
of scalar functions $\uu=[u^l(\xx,t)]^T$, $l=1,2,\ldots,5$, while the set
(\ref{1b}) evolves the vector field $\BB=[B_i(\xx,t)]^T$, $i=x,y,z$.
The overall set of dependent variables are henceforth represented by the
eight-component array $\ww=[\uu,\BB]^T$.
The first array contains the conservative fluid variables
$\uu=[\rho,q_i,e]^T$, where $\rho$ is the mass density, $q_i=\rho v_i$
are the momentum components, $v_i$ are the fluid velocity components,
and $e=p/(\gamma-1)+\rho v^2/2 + B^2/2$ is the total energy density for a
perfect gas equation of state, where $p$ is the kinetic pressure
and $\gamma$ is the adiabatic index.
The corresponding flux vector components $\ff_i=[f^l_i]^T$, $l=1,2,\ldots,5$
are given by $\ff_i=[q_i,M_{i,j},H_i]^T$, $i,j=x,y,z$, with the momentum flux
tensor defined by $M_{i,j}=v_iq_j+\Pi\delta_{i,j}-B_iB_j $
and the energy flux components defined by $H_i=v_i(e+\Pi)-B_i(\vv\cdot\BB)$,
where $\Pi=p+B^2/2$ is the total pressure.
In sub-system (\ref{1b}), which is the induction equation for the magnetic
field vector $\BB$, the corresponding flux is simply given by the electric
field vector $\EE=-\vv\times\BB$, where the assumption of a perfect
conducting plasma (ideal MHD) has been implicitly assumed.

As for the Euler equations, the system (\ref{1a},\ref{1b}) has to
be supplied with entropy functions $S=S(\ww)$ satisfying the condition
\be
{\partial S\over\partial t}+\nabla\cdot\FF_S(\ww)\leq 0,
\label{1d}
\ee
which allows to identify, among discontinuous solutions of the MHD
system, the (physically) admissible ones.
The existence of entropy functions (in fact $S=-\rho s$, where
$s\propto\log(p\rho^{-\gamma})$ is the physical entropy per unit mass)
is also related to the hyperbolic structure of the MHD equations.
For smooth solutions, the system (\ref{1a},\ref{1b}) can be put in the
non-conservative (quasi-linear) form
\be
{\partial\ww \over \partial t}+[\hJJ(\ww)\cdot\nabla]\ww=0,
\label{jacobian}
\ee
where $\hJJ=(\hJ_i)$, $i=x,y,z$, and each $\hJ_i$ is the Jacobian matrix of the
eight-component flux array $[f^l_i,E_i]^T$ with respect to the $\ww$
variables. It is a well known property that any linear combination
$\hJ(\ww,{\bf k})=\sum_ik_i\hJ_i(\ww)$, for real $k_i$ numbers, and then also
each $\hJ_i$ matrix, is hyperbolic at any reference state $\ww$.
Moreover, as for the Euler equations (see \cite{HL}), the (positive) Hessian
matrix $S_{\ww,\ww}$ acts as similarity transform to make all $\hJ_i$
symmetrizable.

To underline differences and analogies
of the MHD system with respect to the reference Euler system which may
have relevance for numerical modeling, some remarks are in order:

\begin{itemize}
\item
The $\uu$ array contains  {\em scalar} variables and
the corresponding flux derivatives
are expressed by the $div\equiv[\nabla\cdot]$ conservative operator.
Sub-system (\ref{1a}) has then the same formal structure of the Euler system
for gas-dynamics. At surface elements where discontinuities take place, this
conservation form leads to the usual Rankine-Hugoniot relations.
On the other hand, the $\BB(\xx,t)$ vector is anti-symmetric (an axial vector),
components $B_i$ are pseudo-scalars, and the corresponding evolution operator
is  given by the anti-symmetric $curl\equiv [\nabla\times\cdot]$ derivative.
The conservative form is now expressed by the scalar condition (\ref{1c})
(magnetic flux conservation) and by the $\nabla\times\EE$ flux derivatives
(conservation along a closed contour).
Discontinuous solutions satisfy jump relations just for the tangential
components $\BB_t=\BB\times\nn$, where $\nn$ indicates the normal
direction, whereas the normal field component $B_n=\BB\cdot\nn$ is
continuous.
The Rankine-Hugoniot relations, once supplied with an appropriate entropy law,
allow to identify the physically correct discontinuous solutions.
It is apparent that magnetic discontinuities and the related entropy
constraint do not involve the parallel $B_n$ component.

\item
It follows that smoothness properties of MHD variables are also
different. Scalar components $u^l(\xx,t)$ may develop discontinuous
solutions along any space direction and can be then
represented on the space of piecewise continuous functions.
The vector field $\BB(\xx,t)$ has more elaborate properties,
since the divergence-free condition entails the $\BB(\xx)$ field
maps piecewise differentiable (and then continuous) field lines.
The conservation law given by Eq.~(\ref{1b}) is then essential to preserve
in time condition (\ref{1c}) and to assure the smoothness properties of the
magnetic field.

\item
The divergence-free condition enters implicitly in the MHD momentum
and energy conservative equations. This can also be expressed by realizing
that the Maxwell tensor $\TT=\II B^2/2-\BB\BB$ in the momentum flux has
to satisfy
\be
\BB\cdot(\nabla\cdot\TT)=0,
\label{1ac}
\ee
in order to recover the correct Lorentz force in non-conservative form.

\item
Finally, the divergence-free condition allows to represent the $\BB(\xx,t)$
field via a vector potential $\AA(\xx,t)$, defined by $\BB=\nabla\times\AA$
and by the gauge condition $\nabla\cdot\AA=0$, which assures the uniqueness
of this representation. The new evolution equation is now
\be
{\partial\AA\over\partial t}+\EE=0.
\label{vecpot}
\ee
The above relations and the induction equation (\ref{1b}), together
with the condition $\EE\cdot\BB=0$ valid for ideal MHD, imply an added
conservation law for the magnetic helicity $H=\int(\AA\cdot\BB){\rm d}x^3$,
carrying informations on the topology of magnetic field lines.
\end{itemize}

When looking at (finite-dimensional) numerical approximations, 
a main problem is that no rigorous results on convergence are available, 
even for the Euler system.
In this case, however, by taking advantage of theoretical achievements,
like the Lax-Wendroff theorem \cite{LW}, {\em heuristic} guidelines are
usually adopted in order to:
\begin{itemize}
\item retain the conservative form of the original equations in the
discretized system;
\item assure consistency, in the sense that the approximations of the flux
functions and of the differential operators have to recover the exact ones
as the spatial and temporal grid sizes go to zero;
\item assure non-oscillatory (or even monotonicity preserving) numerical
representation of discontinuous data;
\item assure consistency with the entropy law, in a way the numerical
viscosity induced by the upwind differentiation is compatible with
Eq.~(\ref{1d}) (see \cite{TA});
\item assure stability of the numerical solution.
\end{itemize}

As already anticipated in the Introduction, the main issue addressed here
is to select a set of {\em additional} requirements
for the MHD system which should assure that the specific properties
of the magnetic field enter as {\em built-in} conditions of a numerical
scheme. We propose the following:
\begin{itemize}
\item
the discretized first derivatives $\partial_iB_i$
entering the $\divb$ definition are consistent approximations;
\item
for initial divergence-free fields the approximated derivatives
satisfy $\divn=0$ exactly;
\item
divergence-free initial conditions are preserved
exactly in time by the discretized induction equation.
\end{itemize}
We then suggest the following definition: {\em a numerical scheme is
consistent with the specific properties of the MHD system} if all above
conditions are fulfilled.
This definition, together with the guidelines for Euler equations,
will enable us to identify and construct a class of Godunov-type
schemes for MHD, later referred to as UCT-based schemes.

In this framework, as for the Euler equations, a finite volume setting
provides a sufficiently general starting point.
Here we concentrate only on algorithms for regular structured grids,
even if the generality of the method allows to extend some basic procedures
also to adaptive mesh refinements (AMR, \cite{BC}) and to unstructured grids.
In particular, De Sterck \cite{DS} has developed a general CT formalism for
unstructured triangular grids, named MUCT, where rigorous geometrical
arguments have been considered to support this approach.

\section{The UCT method to design Godunov-type schemes for MHD}

\subsection{Discretization step: finite-volume formalism}

In a finite-volume setting, the 3-D computational domain $\Omega$ is
first subdivided in Cartesian cells $C$, with volume $V$, side sizes
$h_i$, $i=x,y,z$ and faces given by the oriented surface elements $S_i^{\pm}$,
$i=x,y,z$, where $\pm$ denotes the sign of face normals.
For each face $S_i$, we then denote as $(L^{\pm}_j,L^{\pm}_k)$, $j,k\neq i$
its oriented sides. At this level of the analysis no
indexing on a grid is needed, thus allowing to extend the formalism
also to a non-uniform partition of the $\Omega$ domain, as required in
grid refinement techniques. In the following, a {\em semi-discrete}
finite-volume approach will be  employed, thus only space
averages will be considered and the time dependency will be left for further
integration, for example via standard Runge-Kutta algorithms \cite{SO}.

A conservative discretization of sub-system (\ref{1a}) is accomplished,
as usual for Euler system, by integrating each scalar equation on the volume
element $V$ of each cell $C$. By application of the Gauss theorem, one has
then
\be
\frac{{\rm d}}{{\rm d}t}\auu(t)+\sum_i\frac{1}{h_i}(\aff^+_i-\aff^-_i)=0,
\label{2a}
\ee
where
\be
\auu(t)=\frac{1}{|V|}\int_V\uu(\xx,t){\rm d}V,~~~~~
\aff^\pm_i=\frac{1}{|S_i^\pm|}\int_{S_i^\pm}\ff_i[\ww(\xx,t)]{\rm d}S
\ee
denote respectively volume averages of each scalar component $u^{l}$ over
the cell $C$, and $\aff^{\pm}_i$ are flux values averaged on cell faces
$S_i^{\pm}$. We note that $\aff_i$ fluxes are
represented as exact {\em point values} in the (non averaged) parallel
coordinate $i$, and the corresponding differences in (\ref{2a})
provide the averaged flux derivatives.

In the case of sub-system (\ref{1b}), two different approaches can be
pursued. In the SUP approach, magnetic field components
are discretized by volume averages $\aB_i$ as other
scalar variables $u^l$, and the electric field components by
face averages $\aE_k$, as $\ff_i$ fluxes. We have then
\be
\frac{{\rm d}}{{\rm d}t}\aB_i(t)+\sum_{j,k}\epsilon_{i,j,k}\frac{1}{h_j}
(\aEE^+_k-\aEE^-_k)=0,
\label{2ab}
\ee
where $\epsilon_{i,j,k}$ is the Levi-Civita symbol and $\pm$ here
refers to faces normal to the $j$ direction. 

\begin{figure}
\centerline{\resizebox{.7\hsize}{!}{\includegraphics{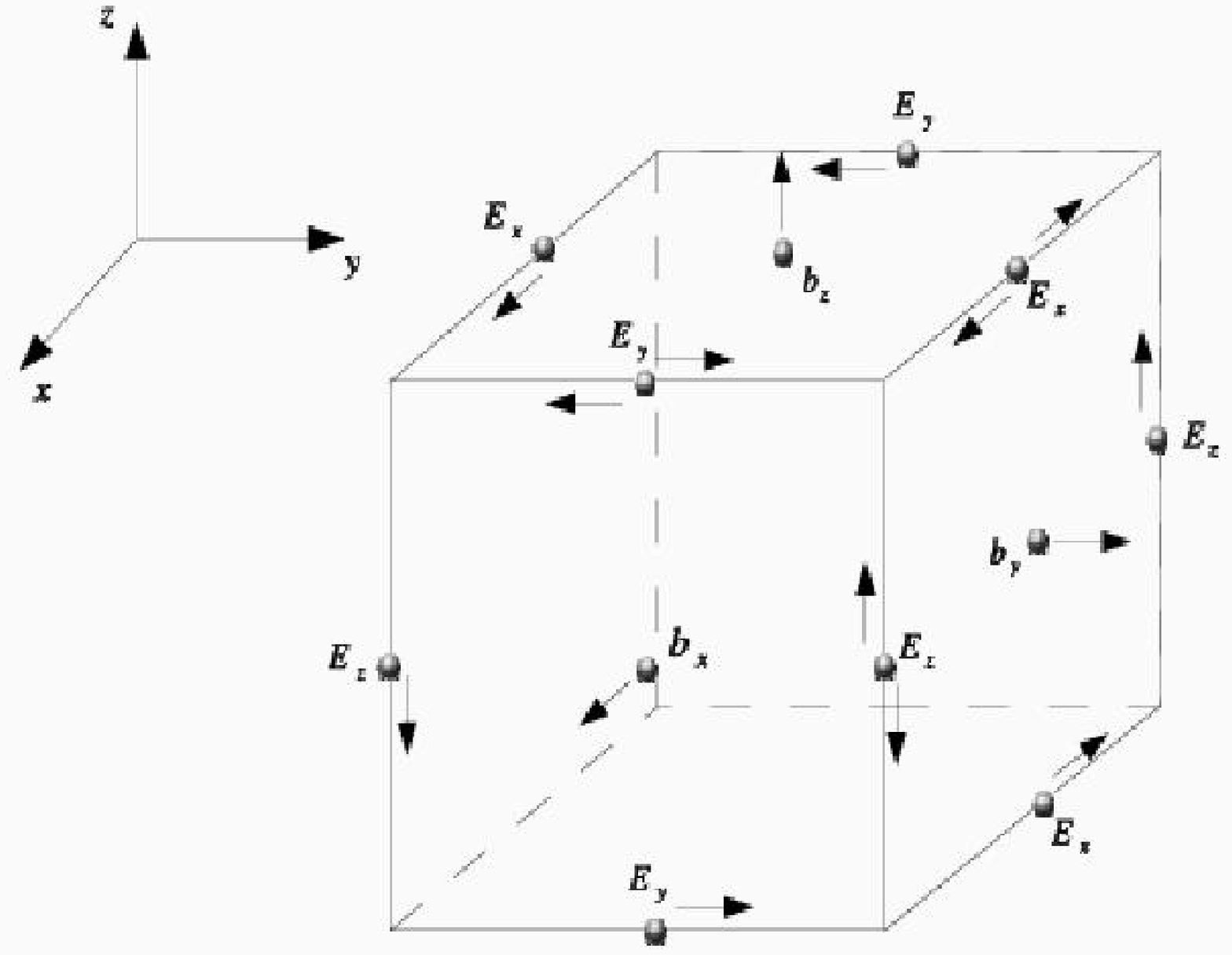}}}
\caption{The staggering of magnetic and electric vector fields in the CT 
framework. Only the $S_x^+$, $S_y^+$ and $S_z^+$ cell faces are visible. 
The arrows indicate the respective face normals, placed at intercell 
centers where $b_i$ magnetic field components are defined, and the 
relative oriented contours for the application of Stokes' theorem, with 
arrows placed at edge centers where electric field components are defined.
}
\label{cube}
\end{figure}

On the other side, in the CT formalism a discretization preserving
the original (vector anti-symmetry) property is accomplished
by a {\em surface} integration on a cell face followed by the
application of the Stokes theorem on the line contour of that face.
This leads to
\be
\frac{{\rm d}}{{\rm d}t}\ab_i(t)+\sum_{j,k}\epsilon_{i,j,k}\frac{1}{h_j}
(\aEE^+_k-\aEE^-_k)=0,
\label{2b}
\ee
where now
\be
\ab_i(t)=\frac{1}{|S_i|}\int_{S_i} B_i(\xx,t){\rm d}S,~~~~~
\aE^\pm_k=\frac{1}{|L_k^\pm|}\int_{L_k^\pm} E_k[\ww(\xx,t)]{\rm d}L
\label{bdef}
\ee
are respectively the {\em staggered} discretized magnetic field variables,
defined as integrals over the cell face $S_i$ (we retain the formalism
of non-capital $b_i$ components to indicate staggered values to conform
with other authors), while $\aE^\pm_k$ are now line-averaged electric field
components along face edges $L_k^\pm$, where the orientation depends on the 
normal to the face under consideration (see Fig.~\ref{cube}).
Here the magnetic field components are represented as (normalized)
{\it magnetic fluxes}, thus $\ab_i$ are exact point values in the parallel 
coordinate $i$ (as $\aff_i$) while $\aE_k$ are point values with respect 
to their orthogonal coordinates $(i,j)\neq k$.

By using this staggered discretization, which yields a couple of normalized
fluxes $\ab_i^\pm$ defined at $S_i^\pm$ faces for each direction $i$, 
it is now possible to represent the volume average of the (parallel) 
first derivatives $\partial_iB_i(\xx)$.
Therefore, for divergence-free initial conditions (\ref{1c}), we have
\be
\sum_i\frac{1}{h_i}(\ab_i^+-\ab_i^-)=0,
\label{2c}
\ee
which will be preserved in time algebraically by the induction
equation (\ref{2b}).

At this general level, anything is exact. Approximations (in space)
arise when all MHD variables, starting from the discretized values,
have to be reconstructed at the cell faces where fluxes
are defined as point values. However, even at this preliminary step,
differences in the magnetic field representations of Eq.~(\ref{2b})
with respect to Eq.~(\ref{2ab}) have relevance:

\begin{itemize}
\item When primary data for the magnetic field are the staggered
$\ab^{\pm}_i$ components, on each cell $C$ one has at disposal
{\em two independent} sets of data. A first consequence is that
{\em no reconstruction} is needed to evaluate (at a second order 
approximation) these variables as argument of the corresponding
$\aff_i$ fluxes at a $S_i$ face. Moreover, these staggered data
carry informations both on the volume averaged (or centered) values
$\aB_i$ and on the first derivative along the parallel coordinate.
In fact, $(\ab^+_i-\ab^-_i)/h_i$ provides a second order 
(and then consistent) approximation of the $\partial_iB_i$ first
derivative in point-wise (or finite difference) sense
{\em inside each cell}.

Related to the above is the property that each $\ab_i$ component
provides a {\em continuous} sampling across the corresponding $S_i$ cell face.
This follows from the definition in Eq.~(\ref{bdef}) and by taking into
account that a divergence-free $\BB(\xx)$ field entails a continuous
elemental flux across a discontinuity surface (see \cite{LD} for details).
In this way, the continuity property of the $B_n$ normal component in the
Rankine-Hugoniot jump relations retains a consistent representation 
in a finite volume setting.

Staggered components can also be defined, in a {\em fully equivalent} way,
by using the vector potential ${\bf A}$. In fact, by face-averaging the 
defining condition $\BB=\nabla\times\AA$, one has
\be
\ab_i=\sum_{j,k}\epsilon_{i,j,k}\frac{1}{h_j}(\aA^+_k-\aA^-_k),
\label{potrep}
\ee
still assuring the divergence-free condition in the form (\ref{2c}).
If needed, a time evolution for the numerical (line averaged)
$\aA_k$ components, consistent with Eq.~(\ref{vecpot}),
can also be formulated:
\be
\frac{{\rm d}}{{\rm d}t}\aA_k(t)+\aE_k=0,
\label{potential}
\ee
preserving in time the representation of Eq.~(\ref{potrep}).
Here, the numerical flux $\aE_k$ are precisely the same as
in Eq.~(\ref{2b}).

\item
On the other hand, when primary data are represented by $\aB_i$
volume averages, as in Eq.~(\ref{2ab}), only {\it one} numerical value 
per cell is available. The averaging procedure, which is now applied 
also along the $i$ coordinate, while still assuring that $\aB_i$ 
are consistent approximations of $B_i$ point values, it also entails a loss 
of {\em direct} information on the point values at the $S_i$ cell faces and
on the correspondent parallel derivatives. Moreover, interpolation
procedures are not sufficient, for discontinuous data, to
recover these informations (related to the divergence-free property)
and then some added argument, procedure, or constraint would be necessary, 
in our opinion, to avoid inconsistent approximations.
In the following section we shall provide a more detailed
analysis on this problem.
\end{itemize}

We can then conclude here that a finite volume CT formulation of the
induction equation satisfy the
the general consistency demands presented in Sect.~1.1, provided the
$\partial_iB_i$ are approximated using staggered $\ab_i$ data.

In spite a CT based discretization for MHD is a longstanding well known
framework for induction equation, its application to the whole MHD system
is yet a matter of debate. In fact, it is yet a persistent viewpoint
in numerical community that staggered collocation of magnetic
field may be useful only to express the divergence-free relation,
being otherwise not well suited for upwind formulation in
Godunov-type schemes. In contrast with this viewpoint,
we propose here to construct and test numerical schemes
explicitly based on the CT discretization and on the related
properties detailed above.

\subsection{Reconstruction step: scalar vs divergence-free fields}

In the following, we specialize to a Cartesian partition of the
computational domain $\Omega$. For grid indexing
$1\leq j\leq N_x$, $1\leq k\leq N_y$, $1\leq m\leq N_z$, a generic
cell $C_{j,k,m}$ is defined as
\be
C\P\equiv [\xm,\xp]\times [\ym,\yp]\times [\zm,\zp],
\ee
where each fractional index labels a cell interface (say $\xp$, here with
$0\leq j\leq N_x$), while a cell center has coordinates $(x_j,y_k,z_m)$, 
with $x_j=(\xm+\xp)/2$, $y_k=(\ym+\yp)/2$, and $z_m=(\zm+\zp)/2$.
For simplicity, now we assume a uniform partition, along all directions,
so that ($h_x\equiv\xp-\xm$, $h_y\equiv\yp-\ym$, $h_z\equiv\zp-\zm$)
are the constant sizes for all cells. Under these settings, the primary 
volume-averaged array of fluid variables will be indicated as $\auu\P$.
Surface-averaged staggered magnetic field components, defined
at cell interfaces, will be indicated as
$(\ab_x)_{\jp,k,m}$, $(\ab_y)_{j,\kp,m}$, and $(\ab_z)_{j,k,\mp}$.
The same notation holds then for face-centered flux components
$\aff_i,\,i=x,y,z$, while the edge-centered $\aE_k$ fields have indexing
$(\aE_z)_{\jp,\kp,m}$, and so forth for other components.
Finally, the divided differences for functions located at cell interfaces
introduced in the previous section, e.g. in Eqs.~(\ref{2a}), (\ref{2b}), 
and (\ref{2c}), will be denoted here as $D_i,\,i=x,y,z$.
For a generic 1-D scalar function $f$ located at inter-cell points $\xpm$,
$D_x$ is then given by
\be
[D_x(f)]_j=\frac{1}{h_x}[\Delta_x f]_j;~~~~ [\Delta_x f]_j=f_\jp-f_\jm.
\label{D}
\ee

In higher order Godunov-type schemes, a one-dimensional scalar variable $u(x)$,
represented by cell-centered data $\{\au_j\}$, is first reconstructed as
approximated point values $\su(x)$ inside any cell $C_j$ and up to the
interior cell faces $\xpm$, where fluxes $f_x(u)$ have to be evaluated.
This is accomplished by using local polynomials $\su(x)$: a) consistent
with the cell averages values $\au_j$; b) having monotone or non-oscillatory 
properties. In a second order approximation, one has the linear fit
\be
\su_j(x)=\au_j+\Dx(\au)(x-x_j),
\label{recu}
\ee
where the non-oscillatory derivative $\Dx(u)$ is usually constructed using
slope limiters. In the simplest case it is defined as
\be
[\Dx(\au)]_j=\frac{1}{h_x}mm([\Delta_x \au]_{j+1/2},[\Delta_x \au]_{j-1/2}),
\label{mm}
\ee
where $[\Delta_x \au]_{l+1/2}=\au_{l+1}-\au_l$ ($l=j,j-1$) and where
$mm(a,b)$ denotes the usual two-point {\em MinMod} (MM) algorithm. More
elaborate limiters can likewise be constructed, using the Van Leer \cite{VL}
monotonicity constraint or TVD \cite{H} properties.
For higher order schemes, ENO-based procedures have been developed
(see \cite{S}, for a review) assuring a non-oscillatory reconstruction
under weaker monotonicity constraints.

The reconstruction in Eq.~(\ref{recu}) extends up to the interior face points
$\xpm$ providing a {\em left approximation} on the
$S^+_x$ face and a {\em right approximation} on the $S^-_x$ face,
along the indicated coordinate, as needed for flux computation.
At a given cell interface, say at $x=\xp$, in cases where a jump of size
$\Delta_x u=O(1)$ occurs, the estimated slope coefficients $[\Dx]_j$
and $[\Dx]_{j+1}$ using minmod limiters both vanish and the reconstructed
$\su(x)$ variable is represented by piecewise constant (first order)
interpolants $\su_j=\au_j$ and $\su_{j+1}=\au_{j+1}$, respectively.

For a multidimensional function $u(\xx)$ (in the uniform grid defined above),
a tensor-product representation with one-dimensional interpolants
on each coordinate is usually adopted. For unstructured grids more elaborate
procedures are required, but the basic ingredients (consistency with the
cell averaged data and non-oscillatory constraints) still hold.
In regular grids, the resulting second order approximation
$\su(\xx)\approx u(\xx)$ on each cell $C\P$ takes then the form
\be
\su(\xx)=\au+\Dx(\au)(x-x_j)+\Dy(\au)(y-y_k)+\Dz(\au)(z-z_m),
\label{rec3}
\ee
where all quantities are implicitly calculated at cell center and each
$\sD_i(\au)$, is the non-oscillatory 1-D first derivative defined in
Eq.~(\ref{mm}).

When dealing with the magnetic field $\BB(\xx)$, a different approach is
needed, in general, to take into account the vector structure and
the specific smoothness properties
already quoted in the previous sections. Starting with face-averaged data
$\ab_i$, for $i=x,y,z$, satisfying the divergence-free condition (\ref{2c})
\be
D_x(\abx)+D_y(\aby)+D_z(\abz)=0,
\label{divb}
\ee
the problem of representing each $B_i(\xx)$ field along the proper parallel
coordinate inside a cell is already solved at the linear level, since the 
slope $D_i(\ab_i)$ is at disposal. Instead, reconstruction is needed along
the face (orthogonal) coordinates, where the field is sampled by averaged
$[\ab_x(x)]_{k,m}$ data.

For the $B_x(\xx)$ function one has then
\be
\sB_x(\xx)=\sB_x +D_x(\abx)(x-x_j)+\Dy(\sB_x)(y-y_k)+\Dz(\sB_x)(z-z_m).
\label{recbx}
\ee
Here, the cell-centered $\sB_x$ values result by noticing
that $\sB_x(x,y_k,z_m)$ is
the the unique linear interpolant of a continuous function
with data at the $\xpm$ points. We have then

\be
(\sB_x)\P=\frac{1}{2}[(\abx)_\jp+(\abx)_\jm]_{k,m},
\label{abx}
\ee
which provides an {\em approximation} of the cell averages and of the
$B_x$ point values:
\be
\sB_x=\aB_x+O(h_x^2)=B_x+O(h_x^2).
\ee

In Eq.~(\ref{recbx}) it is evident how staggered and cell-centered fields
work differently. Along the parallel $x$ coordinate, the $\sB_x(x)$
function is entirely defined by the $(\abx)_{\jpm,k,m}$ data, providing
the field values at the cell interfaces, the approximated first derivative
$[D_x(\abx)]\P$, and the approximated cell-averaged value $(\sB_x)\P$.
Reconstruction is needed instead to evaluate the $B_x$
left-right values at the orthogonal cell faces $(S^{\pm}_y,S^{\pm}_z)$,
where $B_x$ may have discontinuities and where, correspondingly,
$\ab_x$ and cell-centered values $\sB_x$ behave as the other
$\au$ scalar variables.

A second property related to the divergence-free conditions
is that the $\sB_x(x,\cdot)$ function in Eq.~(\ref{recbx})
maps a continuous function with first derivative
$D_x (\ab_x)$ which may be discontinuous. To evaluate the jump size,
one considers the second difference
$\Delta^2_x\abx\equiv\Delta_x\Delta_x\abx$ centered at the proper
interface point $(\xp,y_k,z_k)$. By taking into account the divergence-free
relation (\ref{divb}) and the commutativity property of the two-point
difference operators, one has
\be
{1\over h_x}\Delta^2_x\abx=-{1\over h_y}\Delta_y\Delta_x\ab_y-
{1\over h_z}\Delta_z\Delta_x\ab_z,
\label{dem}
\ee
where at least one of the differences $\Delta_x\ab_y$ or $\Delta_x\ab_z$
has size $O(1)$, by definition.

Following the same procedure the $(B_y,B_z)$ components can be represented,
on the same cell, by the relations
\be
\sB_y(\xx)=\sB_y+\Dx(\sB_y)(x-x_j)+D_y(\aby)(y-y_k)+\Dz(\sB_y)(z-z_m),
\label{recby}
\ee
\be
\sB_z(\xx)=\sB_z+\Dx(\sB_z)(x-x_j)+\Dy(\sB_z)(y-y_k)+D_z(\abz)(z-z_m),
\label{recbz}
\ee
and remarks made above on different behaviors of staggered $(\aby,\abz)$
and cell-centered $(\sB_y,\sB_z)$ values, respectively, and on the different
smoothness properties depending on the involved coordinates, apply.

We remark here that this apparent {\em duality} in the reconstruction 
procedure appears to be fully consistent with the physical
duality of the MHD Rankine-Hugoniot relations. In fact, using a local
characteristic decomposition of the $(\uu,\BB)$ variables, for a specified
direction, say $x$, only the $(\auu,\sB_y,\sB_z)$ variables participate to
the Riemann wave fan and may contribute, then, to the discontinuous
characteristic modes. On the other hand, the continuous $\ab_x$ variable
has no role in the related upwinding procedures.

In upwind schemes, the reconstruction step presented above
has relevance not only to recover face centered
values, but also to approximate the $\BB(\xx)$ field at any point inside
a cell, as it is required in schemes adopting grid refinements (AMR) or
multi-grid procedures. In this context, divergence-free interpolants
similar to those derived here, even if based on quite different arguments,
have been proposed in \cite{B}. 
A related work \cite{TR} has presented a detailed analysis to show that, 
under linear interpolation based on staggered data, conservative properties
and the divergence-free relation can be preserved in cell-subcells
refinement procedures.

To summarize,
it follows from this analysis that a numerical divergence-free magnetic field
can be represented in an unambiguous way by using $\ab_i$ staggered values
(or equivalently the related numerical vector potential) as primary data.
In this representation, second order approximated  first derivatives are
consistent and non-oscillatory (no cell crossing is needed) and the variable 
$\divn\equiv\sum_i [D_i(\ab_i)]\P$ in Eq.~(\ref{divb}) results to be exactly
zero inside any point of $C\P$. At the same time, the reconstruction
procedures at cell interfaces, where variables are discontinuous,
provide definite rules on how to formulate upwind differentiation.

\subsubsection{Reconstruction and central derivatives in non-CT schemes}

A CT-based formalism helps also to analyze the reconstruction problem in 
the SUP framework, where only cell averaged $\aB_i$ values are at disposal.
By restricting to the $B_x(x,\cdot)$ function, a linear interpolant reads
\be
\sB_x(x,\cdot)=[\aB_x]_j +[C_x(\aB_x)]_j(x-x_j)+..
\label{recBx}
\ee
where $C_x(\aB_x)$ denotes a (by now unspecified) {\em consistent}
approximation of the first derivative. For a piecewise differentiable
function, at a point $\xp$ this interpolant provides in general {\em two}
values, as left and right approximations, like for all other variables. 
By imposing there the additional continuity condition, specific to
magnetic field components, one has
\be
[\aB_x]_j +{1\over 2}h_x [C_x(\aB_x)]_j=
[\aB_x]_{j+1} -{1\over 2}h_x [C_x(\aB_x)]_{j+1},
\label{derBx}
\ee
and since Taylor expansion for $C_x$ is not applicable across a discontinuous 
interface, this assures only an {\em implicit} way to express
the numerical derivative in terms of the cell centered data.
This suggests that some added condition, like the non-oscillatory constraint,
should be required even for $x$-wise interpolations, but how to recover 
the divergence-free condition remains here an open question.

We are not aware of any SUP-based scheme where this problem
has been properly addressed. It is a common practice, instead, to
use for flux computations the mid-point average
\be
(\sb_x)_{\jp,k,m}=\frac{1}{2}[(\aB_x)_{j,k,m}+(\aB_x)_{j+1,k,m}],
\label{midpoint}
\ee
resulting in a $O(h_x)$ approximation. In turn, this entails an approximation 
for the first derivative given by the {\em central difference}
\be
[D_x^{(c)}(\aB_x)]\P={1\over 2h_x}[(\aB_x)_{j+1,k,m}-(\aB_x)_{j-1,k,m}].
\label{Dc}
\ee
However, when the relevant $(x_{j-1},x_{j+1})$ stencil includes a 
discontinuity, say at $\xp$, by using the relation (\ref{derBx}) with 
$C_x=D_x(\ab_x)$ we have
\be
[D_x^{(c)}(\aB_x)]\P=[D_x(\ab_x)]\P +O(1),
\ee
the $O(1)$ term resulting from the first derivative jump as
estimated in (\ref{dem}).
Using the same argument to the other $(B_y,B_z)$ components,
a final  $\divn=O(1)$ results. This is well documented
in numerical experience where a central difference
is used for discontinuous functions (the Gibbs pathology).

\subsubsection{Extension to higher order}

The first derivative discontinuity of a divergence-free field has also 
relevance to extend a CT-based reconstruction to higher orders ($r\geq 3$)
of spatial accuracy. In Godunov-type schemes, higher order reconstructions
for scalar variables are usually provided by ENO-based interpolants, 
like Weighted-ENO (WENO, \cite{S} and references therein) and Convex-ENO
(CENO, \cite{LO}). For regular grids, these interpolants, once defined for
one-dimensional variables, can be extended to higher dimensions by
a tensor-product representation.

For the magnetic field more elaborate procedures are
needed, however, by taking into account that non-oscillatory
derivatives along different directions are non-commutative.
A general strategy we propose is to reconstruct first the
vector potential components $\aA_i$ in the usual way, by taking advantage
that these are {\em scalar} variables, and then to define the $b_i$
point values using the basic $\BB=\nabla\times\AA$ relation. However,
using the vector potential alone is not sufficient to guarantee
a divergence-free relation. A crucial step to make this procedure
effective is to approximate the $\nabla\times\AA$ derivatives by 
consistent two-point, fixed-stencil, high order finite differences.

As an example, in the Appendix we report the third order implementation
of the CENO reconstruction procedures.

\subsection{Upwind step: Roe-type approximate Riemann solvers}

Using the grid notation of the Sect.~2.2, the MHD Eqs.~(\ref{2a}) and
(\ref{2b}) take on the form
\be
\frac{{\rm d}}{{\rm d}t}[\auu(t)]+\sum_i D_i(\aff_i)=0,~~~~
\label{3a}
\ee
and
\be
\frac{{\rm d}}{{\rm d}t}[\ab_i(t)]+
\sum_{j,k}\epsilon_{i,j,k} D_j(\aE_k)=0,
\label{3b}
\ee
the first set being centered at cell nodes and the second at cell interfaces.
The overall system has now to be evaluated using some approximate Riemann
solver, the same for all flux functions and for all Cartesian components,
at a time. Here we make reference to Roe-type schemes, allowing a full
resolution of the characteristic MHD modes, whereas the so-called central
and central-upwind schemes, which avoid spectral decomposition, will be
briefly treated in Sect.~3.2

\begin{figure}
\centerline{\resizebox{.7\hsize}{!}{\includegraphics{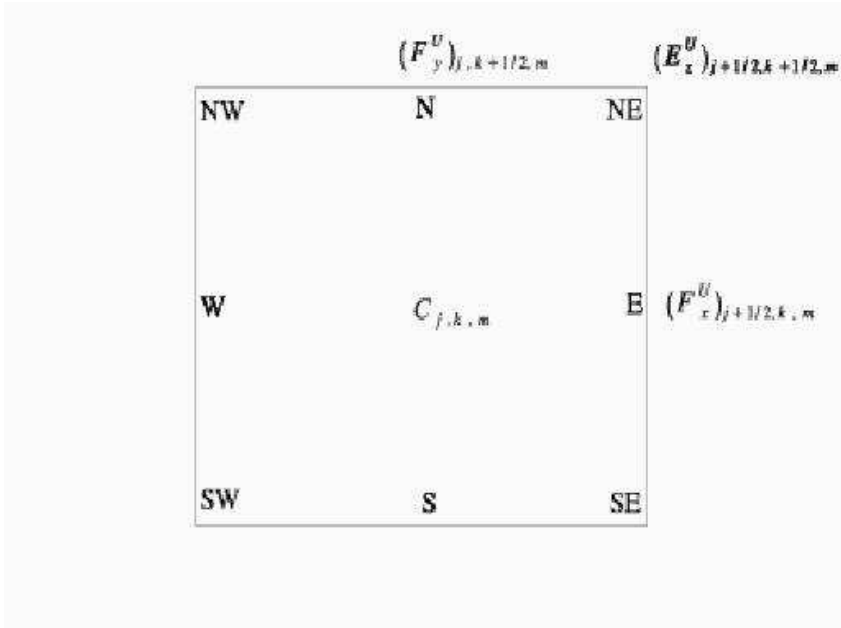}}}
\caption{The notation used for the reconstructed values inside
a cell $C\P$ (here a cut through the center, normal to the $z$ 
direction, is shown) and the position of upwind fluxes, to be 
constructed via contributions from neighbouring cells.
Fluxes are either defined as two-state functions located at cell 
interfaces, $(\FF^U_x)_{\jp,k,m}$ and $(\FF^U_y)_{j,\kp,m}$, 
or as four-state functions located at cell edges, $(\EE^U_z)_{\jp,\kp,m}$. 
}
\label{upwind}
\end{figure}

Let us then first specialize to the 1-D flux differentiation along the $x$
coordinate and denote with $\FF_x=[F^s_x]^T$, for $s=1,2,\ldots,7$, the array 
of all flux components, defined at the $x=\xp$ point for generic $(y,z)$
coordinates of the $S_x=S_x^+$ face. These components are
$F_x^{(l)}=\fl_x$, for $l=1,2,\ldots,5$, whereas the components entering
the induction equation are $F_x^{(6)}=-E_z$ and $F_x^{(7)}=E_y$.
Correspondingly, we denote with $\ww_x=[\rho,q_i,e,B_y,B_z]^T$ the array 
of variables which need reconstruction as point values at the $S_x$ face and
as $[\ww_x^E(y,z),\ww_x^W(y,z)]$ (East-West, see Fig.~\ref{upwind}) 
the corresponding left-right states. The magnetic field component 
$b_x(x,\cdot)$ in the parallel direction $i$ satisfies $b_x^E=b_x^W$ 
by continuity, thus the arguments of flux functions $\FF_x(\ww)$ have then 
to be specified as $(\ww_x^a,b_x)$, for $a=E,W$.
For short, we denote as $\FF_x^a=\FF_x(\ww_x^a,b_x)$.

Approximated Riemann solvers based on local linearization technique
(see e.g. \cite{CG}, \cite{MR} for the MHD case) rely on the Roe matrix
$\hA_x$, defined as numerical Jacobian by
\be
\FF_x^W-\FF_x^E=\hA_x(\sww)\cdot(\ww_x^W-\ww_x^E)
\equiv\hA_x\cdot\delta_x\ww_x,
\label{df}
\ee
at any $(y,z)$ point of the indicated $S_x$ cell face.
As usual, the $\hA_x$ matrix is evaluated at an appropriate
intermediate state $\sww=[\sww_x(\ww_x^W,\ww_x^E),b_x]$,
and is required to be consistent with the Jacobian matrix $\hJ_x$ presented
in Sect.~1.1. We notice the $\hA_x$ matrix has rank {\em seven}, as in the
pure 1-D case where $B_x={\rm const}$ holds. In fact, in the
multidimensional case the continuity condition $\delta_x b_x=0$
plays a similar role, implying $B_x$ is locally constant and does not
participate to the characteristic wave fan.

To express the local flux variations in Eq.~(\ref{df}) in terms of
characteristic modes, let us consider the spectral decomposition
$\hA_x=[R\Lambda R^{-1}]_x$, where, if $\lambda_s,s=1,2,\ldots,7$, are
the Roe matrix real eigenvalues, $\Lambda=\mathrm{diag}\{\lambda_s\}$,
and columns of the $R$ matrix are the corresponding right eigenvectors.
In this representation, let us then split
$\hA_x=[\hA_x]^+ +[\hA_x]^-$, where the first term
contains components with $\lambda_s >0$ and the second
with $\lambda_s <0$, respectively. In this form, one has also
$|\hA_x|=[\hA_x]^+ -[\hA_x]^-$, where $|\hA_x|=[R |\Lambda| R^{-1}]_x$
and $|\Lambda|=\mathrm{diag}\{|\lambda_s|\}$.
By using standard procedures, from the two-point flux function
$\FF_x(\ww_x^E,\ww_x^W,b_x)$ a one-valued, continuous and monotone
flux can be selected as upwind state by
\be
\FF^U_x=\FF_x^E+[\hA_x]^-\cdot\delta_x\ww_x=
\FF_x^W-[\hA_x]^+\cdot\delta_x\ww_x,
\label{SPLIT}
\ee
so that the usual Roe flux formula comes out as
\be
\FF^U_x(\ww_x^E,\ww_x^W,b_x)=\FFs_x-\PPhi_x,
\label{ROE}
\ee
where
\be
\FFs_x=\frac{1}{2}(\FF_x^W+\FF_x^E);~~~~~~
\PPhi_x=\frac{1}{2}|\hA_x|\cdot (\ww_x^W-\ww_x^E),
\ee
the first term expressing the smooth component leading to a centered
two-point formula in flux differentiation and the second the Roe-type
component coming from the upwind procedure.
At a discontinuity interface, the latter provides
numerical dissipation which needs to be consistent with entropy conditions.
To that purpose, for shock solutions where $\lambda_s\simeq 0$, 
a small amount of added dissipation $\lambda_s\to\lambda_s+\eta_s$ 
has to be introduced as entropy-fix to avoid unphysical behaviors.

With a straightforward extension, for coordinates $i=y,z$,
the $\hA_i$ Roe matrices and the upwind fluxes
$\FF_i^U(\ww_i,b_i)=\FFs_i-\PPhi_i$ can likewise be constructed, each 
function being evaluated at points of the $S_i$ proper orthogonal face.

By taking advantage of the Roe formalism, which is based on 
{\em independent} 1-D $\hA_i$ matrices, where characteristic modes are
represented locally as planar waves along each direction, it is possible
to evaluate, at a time, numerical fluxes collocated at different points.
In fact, in Eqs.~(\ref{3a}) the final five-component numerical fluxes
$\aff_i=\aFF_i^{(1-5)}$ for $i=x,y,z$ are obtained as an average over the 
proper face $S_i$. This average involves only interior face points and then 
only a characteristic wave fan, the one represented by the $\hA_i$ matrix.
On the other hand, the remaining $\FF_i^{(6-7)}$ flux components appearing in
the induction equation (\ref{3b}) are defined as {\em point values} at the 
intersections of cell faces, where {\em different} characteristic wave fans 
overlap. These flux components can be likewise evaluated by a 
linear combination of 1-D upwind fluxes along the intersecting direction.
It is a main feature of the UCT method that this combination follows a
proper upwind selection rule, since a {\em same flux component} at
the {\em same collocation point} results to have {\em two independent}
representations in terms of characteristic wave fans.

As a prototype, we consider the $E_z$ flux, which is defined at the
$(\xp,\yp,z)$ points where faces $S_x$ and $S_y$ intersect. Let then
denote as $\ww_y=[\rho,q_i,e,B_x,B_z]^T$ the set of variables having a
representation in terms of the $\hA_y$ matrix eigenmodes and
by $\FF_y=\FF_y(\ww_y^S,\ww_x^N,b_y)$, the corresponding flux array, 
where now $(\ww_y^N,\ww_y^S)$ (North-South) denote the left-right states 
along the $y$ coordinate. At the indicated intersection points,
$E_z=-F_x^{(6)}=F_y^{(6)}$ comes out to be a {\em four-state} 
function $E_z(\ww^{a,b})$ where $a=N,S,\,b=E,W$, since the $\ww$ argument 
contains both $\ww_x$ and $\ww_y$ variables (see again Fig.~\ref{upwind}).

The sixth flux component $F^{(6)}_x=-E_z$, defined in Eq.~(\ref{ROE})
and specialized at the $y=\yp$ point, is represented by two independent 
contributions coming from states $(\ww^N,\ww^S)$:
\be
[F_x^U(\ww^N)]^{(6)}={\fs_x}^N-\phi_x^N,\quad
[F_x^U(\ww^S)]^{(6)}={\fs_x}^S-\phi_x^S,
\ee
where $\fs_x={\Fs_x}^{(6)}$ and $\phi_x=\Phi_x^{(6)}$.
On the other hand, the $F^{(6)}_y=E_z$ flux function, defined for generic
$x$ values, is represented by two independent contributions for states
$(\ww^E,\ww^W)$ at the $x=\xp$ point:
\be
[F_y^U(\ww^E)]^{(6)}={\fs_y}^E-\phi_y^E,\quad
[F_y^U(\ww^W)]^{(6)}={\fs_y}^W-\phi_y^W,
\ee
where now $\fs_y={\Fs_y}^{(6)}$ and $\phi_y=\Phi_y^{(6)}$.
By taking into account that $F^{(6)}_y=-F^{(6)}_x$ at the same $(\xp,\yp)$
point, a one-valued numerical flux function having the continuity and upwind
properties along each direction can be then constructed at the
linear level by
\be
E_z^U(\ww)=\Es_z-\phi_y+\phi_x,
\label{EzROE}
\ee
where
\be
\Es_z={1\over 4}[E_z^{NE}+E_z^{SE}+E_z^{NW}+E_z^{SW}];~~~~
\phi_y={1\over 2}(\phi_y^E+\phi_y^W),~~
\phi_x={1\over 2}(\phi_x^N+\phi_x^S),
\ee
and $E_z^{a,b}=E_z(\ww^{a,b})$, with $a=N,S,\,b=E,W$. The continuity
property of this relation implies, in particular, that for any possible
orientation, namely along a cell face diagonal or along a cell face side,
the corresponding one-dimensional planar mode is taken into account
in proper way, to within $O(|\delta_x\ww_x|\, |\delta_y\ww_y|)$,
by the upwind combination above.
The $E_z$ flux, given in Eq.~(\ref{EzROE}) for
generic $z$ coordinate of the $(S_x,S_y)$ common side, is finally
evaluated by line averaging as $(\aE_z)_{\jp,\kp,m}$ numerical flux.
In a similar way, the other $x$ and $y$ components can then
be constructed by a proper combination of upwind fluxes
along the corresponding orthogonal coordinates and line
averaging along the parallel coordinate.

This completes the presentation of the UCT formalism. For later
reference, we quote here the flux formulas given above, specialized
to a first order approximation, which constitutes the building block of
any Godunov-type scheme, needed for higher order extensions.
A piecewise constant reconstruction
for $\ww_x$ variables at the $x=\xp$ point, needed in Eq.~(\ref{ROE}), gives
$\ww_x^E=(\ww_x)_j,\,\ww_x^W=(\ww_x)_{j+1}$, at any $(y,z)$ point.
Face averaging reduces to the one-point evaluation
$\aFF_x(\ww_x,b_x)=\FF_x(\aww_x,\abx)$ and the first order numerical flux
in Eq.~(\ref{ROE}), for the five-component $\ff$ flux array, reads then
(over-bars are henceforth omitted for brevity):
\be
(\ff_x^U)_{j+1/2,k,m}=(\ffs_x)_{j+1/2,k,m}-
(\PPhi_x)^{(1-5)}_{j+1/2,k,m},
\ee
where, at the indicated $(y_k,z_m)$ points,
\be
(\ffs_x)_{j+1/2}={1\over 2}
[\ff_x((\ww_x)_{j+1},(b_x)_{j+1/2})+\ff_x((\ww_x)_j,(b_x)_{j+1/2})],
\ee
and
\be
[\PPhi_x]_{j+1/2}={1\over 2}|A_x(\sw_{j+1/2})|\cdot
[(\ww_x)_{j+1}-(\ww_x)_j],
\ee
and so forth for other directions.
To the same first order approximation, the $E_z$ flux in Eq.~(\ref{EzROE})
can be calculated at $(\xp,\yp,z_m)$ edge points (the $z_m$ centering will
be assumed implicitly in the following). By specializing the
$E_z=\fs_y=-\fs_x=-(v_xB_y-v_yB_x)$ argument variables,
the smooth term results as:
\be
(\Es_z)_{\jp,\kp}=-{1\over 2}[(\hv_xb_y)_{j+1}+(\hv_xb_y)_j]_\kp
                  +{1\over 2}[(\hv_yb_x)_{k+1}+(\hv_yb_x)_k]_\jp,
\label{Ezfirst}
\ee
where $(\hv_x)_{l,\kp}={1\over 2}[(v_x)_k+(v_x)_{k+1}]_l$, for $l=j,j+1$,
and $(\hv_y)_{\jp,l}={1\over 2}[(v_y)_j+(v_x)_{j+1}]_l$, for $l=k,k+1$.
It is worth noticing that in Eq.~(\ref{Ezfirst}), the $\Es_z$ term contains
the $(b_x,b_y)$ staggered components and the resulting four-states
combination at the $(\xp,\yp)$ point {\em cannot} be reduced simply to an
interpolation or averaging form based on the four cell centered
values of the argument.
On the other hand, in the dissipative Roe-type fluxes,
centering at $y=\yp$ of the $\phi_x$ term comes out as
a two-point average in the orthogonal coordinate and correspondingly for
the $\phi_y$ term, so that
\be
(\phi_x)_{j+1/2,k+1/2}={1\over 2}[(\phi_x)_k+(\phi_x)_{k+1}]_\jp,\quad
(\phi_y)_{j+1/2,k+1/2}={1\over 2}[(\phi_y)_j+(\phi_y)_{j+1}]_\kp.
\ee
On the computational side, this form is also economical and of easy
implementation, since the four contributes there involved can be derived from
the $\PPhi_x$ and $\PPhi_y$ fluxes already worked out for the fluid variables.

\subsection{On the problem of numerical monopoles}

We consider now some main differences of the present approach
with other schemes, by focusing on the problem of numerical monopoles.
These unwanted effects may arise when the $\sum_i D_i(\aff_i)$ term
for momentum equations in Eq.~(\ref{3a}) fails to recover
the proper $[{\bf J}\times\BB]_{num}$ Lorentz force in the original
non-conservative form, or, equivalently, when the orthogonality
condition (\ref{1ac}) is not satisfied with sufficiently high accuracy.

This problem has been analyzed in details by \cite{T2}, in a general
discretization setting and with no particular reference to specific upwind
differentiations. Here we follow some of his arguments and notations
to show the behavior of various classes of MHD schemes in comparison
with our UCT method. We specialize to second order flux differentiation,
as it is implemented in most of the schemes in the literature.

By restricting for sake of simplicity to a two-dimensional configuration,
the Maxwell stresses $T_{i,j},\, i,j=x,y,$ entering the momentum flux
components are given by
\be
T_{x,x}=-T_{y,y}={1\over 2}(B^2_y-B^2_x),\quad T_{x,y}=T_{y,x}=-B_xB_y,
\label{mxw}
\ee
evaluated at the proper interface points, that is $(\xpm,y_k)$ for
$T_{x,x}$ and $T_{y,x}$, $(x_j,\ypm)$ for $T_{y,y}$ and $T_{x,y}$.
We notice that these terms refer specifically to the $\fs_i$ smooth
part of the relevant flux components, since Roe-matrix contributions
give only a diffusive term of the Lorentz force. 
For flux differences related to the $q_x$ momentum, one has then the
algebraic relations
\be
D_x(T_{x,x})= \mu_x(B_y)D_x(B_y)-\mu_x(B_x)D_x(B_x),\quad
D_y(T_{y,x})=-\mu_y(B_y)D_y(B_x)-\mu_y(B_x)D_y(B_y),
\ee
all being centered on a $(x_j,y_k)$ point, where $D_x$ and $D_y$
are the usual divided differences of Eq.~(\ref{D}), and where
for components $a=B_x,B_y$ we define the averages
$\mu_x(a)=(a_{j+1/2,k}+a_{j-1/2,k})/2$ and
$\mu_y(a)=(a_{j,k+1/2}+a_{j,k-1/2})/2$.
These two-point averaging on the cell center give
$\mu_x(a)=a_{j,k}+O(h_x^2)$ and $\mu_y(a)=a_{j,k}+O(h_y^2)$,
for second order numerical fluxes.
By summing the two differences defined above, one has (let $h=h_x=h_y$)
\be
D_x(T_{x,x})+D_y(T_{y,x})=-B_x[D_x(B_x)+D_y(B_y)]+
B_y[D_x(B_y)-D_y(B_x)]+O(h^2),
\ee
where the second term on the right hand side provides the numerical
approximation of the $x$ component of the Lorentz force.
By considering then the correspondent $q_y$ momentum flux, the approximation
to the orthogonality condition finally becomes:
\be
[\BB\cdot (\nabla\cdot\TT)]_{num}=-(B_x^2+B_y^2)\divn + O(h^2),
\label{lorentz}
\ee
which is satisfied to within the truncation error if
the numerical divergence-free relation holds (at least) to the
same accuracy order. For smooth flows this is clearly true in any
discretization of first derivatives. But when discontinuities are
present, this term can be even of $O(1)$ size.

In particular:

\begin{itemize}
\item
In schemes where $(B_x)_{j+1/2,k}$ and $(B_y)_{j,k+1/2}$ are reconstructed 
as two-point average of the corresponding cell-centered data, as discussed 
in Sect.~2.2.1, the resulting $\divn$ in Eq.~(\ref{lorentz}) is expressed
by the central differences $D_x^{(c)}(B_i)$, giving contributions
of $O(h^2)$ size in smooth regions and of $O(1)$ size near discontinuous
interfaces.

In Powell's eight-wave scheme, the source terms introduced
in the momentum equation given by $\BB (\divn)$ 
yield a subtraction of these monopoles, which 
is another way to express the Lorentz force in its non-conservative form.

In projection schemes, a $\divn=0$ condition, yet based on cell
centered data, is enforced at each time step as an added constraint,
so that monopoles, which arise when flux derivatives are computed,
are prevented to grow in time.

In previous CT-based schemes (e.g. \cite{DW}, \cite{RMJF}, \cite{BS}),
where staggered variables are actually at disposal, only cell-centered
averaged data $\sB_i$, derived as in Eq.~(\ref{abx}), are then employed
in flux computations.
In that case the resulting $\divn$ term in Eq.~(\ref{lorentz})
is still expressed via derivatives based on central differencing
as in Eq.~(\ref{Dc}), and numerical monopoles are now produced.
These can be evaluated now exactly by noticing that
\be
D_x^{(c)}(\sB_x)=D_x(\ab_x)+h_x^2R_x,\quad
R_x={1\over 4h_x^3}\Delta^3_x \ab_x
\ee
and by similar expressions for the other $(B_y,B_z)$ components.
In this relation the residual $h_x^2R_x$ is $O(1)$ at points where the
$\ab_x$ first derivative is discontinuous, as can be seen from
Eq.~(\ref{dem}). The final estimate gives $\divn=\sum_i h^2_iR_i=O(1)$
since the residuals $R_i,\, i=x,y,z$, being related to the transverse jumps,
do not cancel out, in general.

\item The class of UCT schemes proposed here prevents the onset of monopoles,
since $\ab_i$ staggered data enter directly in the corresponding $\fs_i$
flux components, and $\divnn=\sum_i[D_i(\ab_i)]=0$ results in
Eq.~(\ref{lorentz}).
\end{itemize}

We finally notice that, in order to avoid numerical monopoles,
flux derivatives have to be computed at the {\em same} time-stepping
level, since time-splitting techniques prevent exact cancellation of
$\divn$ terms.

\section{Examples of application of the UCT method}

In the present section, the numerical strategies outlined in Sect.~2
will be applied to a couple of existing Godunov-type schemes, originally
designed for fluid dynamics. We have chosen two completely different
schemes, a classical Roe-type scheme based on field-by-field limiting
along characteristics, and a simple central-type scheme adopting a two-speed
upwind flux with component-wise limiting in the reconstruction algorithms.
Both schemes are proposed here in the semi-discrete form, appropriate
for our UCT method, and then integrated using TVD Runge-Kutta
time discretizations of the appropriate order \cite{SO}.

\subsection {Roe-type: the positive scheme}

We present here the MHD implementation of a second order flux-limited
scheme proposed by Liu and Lax (\cite{LL1}, \cite{LL2}).
This approach allows for an easy formulation for multidimensional hyperbolic
systems and can then represent a well behaved alternative to standard
TVD-based schemes (see \cite{H}). Moreover, these authors have introduced a
new {\em positivity} principle, which is more appropriate
for multidimensional systems, to which TVD do not apply. This stability 
principle relies, in particular, on the symmetrizable form of the system
under investigation and is then well suited also for MHD equations.

Flux-limited schemes are constructed as a
proper combination of an accurate {\em second order} smooth numerical flux,
for example the centered approximation $\FFs=(\FF_{j}+\FF_{j+1})/2$
or a Lax-Wendroff term, and a {\em first order} diffusive flux of the form
$\FF^{diss}=\FFs-\PPhi^{(1)}$, in a way that when the flow is smooth
$\FF^U=\FFs$ and when discontinuities are present $\FF^U=\FF^{diss}$.
In a symbolic way, this combination can be expressed as 
$\FF^U=\FFs-(I-L)\PPhi^{(1)}$, where $L$ is a diagonal operator
whose entries are {\em flux limiter} functions $\phi_s(\theta_s)$ acting,
in general, on characteristic modes and assuring
$\phi_s(\theta_s)=1$ for smooth modes and $\phi_s(\theta_s)=0$ otherwise.

In particular, in this {\em positive} scheme by Liu and Lax, {\em two}
first order dissipative schemes are combined with different flux limiters:
a first (least dissipative) flux of Roe-type and a second (more dissipative)
flux of LLF-type, the latter acting also as entropy-fix for the former.
In the MHD implementation, the Roe matrices $\hA_i(\sww)$ are constructed
in the space of primitive variables, with a simple two-point average
to define the state $\sww$ and with the eigenvector formulation given
in \cite{RB} to remove degeneracies.

The resulting formula, for each cartesian flux component, reads then: 
\be
\FF^U_\jp=\frac{1}{2}[\FF(\ww_{j})+\FF(\ww_{j+1})]-
\frac{1}{2}R[c_{\rm ROE}\,|\Lambda|(I-L)+c_{\rm LLF}\,\alpha (I-M)]
R^{-1}(\ww_{j+1}-\ww_j),
\label{pos}
\ee
where the two coefficients combining the two dissipative terms 
are chosen such as $c_{\rm ROE}+c_{\rm LLF}= 1$, $M$ is a diagonal
term made up of smooth minmod-type limiters that satisfy 
$0\leq\phi_s(\theta_s)\leq 1$, $0\leq\phi_s(\theta_s)/\theta_s\leq 1$, 
$L$ is a diagonal term whose sharper limiters satisfy 
$0\leq\phi_s(\theta_s)\leq 2$, $0\leq\phi_s(\theta_s)/\theta_s\leq 2$, 
for a resulting maximum CFL number of $0.5$ (for more technical aspects, 
the reader is referred to the cited works).
 
In our experience, when applied to the MHD system by following the UCT
strategies described in Sect.~2, this scheme results to be a robust and 
accurate flux-limited scheme, with almost no extra computational effort,
if compared to standard multidimensional TVD schemes.

\subsection{Central-type schemes}

In the fluid-dynamics community, schemes adopting simple one or two-speed
numerical fluxes with component-wise reconstruction (no characteristic 
decomposition is thus required), are often referred to as central or
central-upwind schemes, since it has been proved these approximate Riemann 
solvers come out in Roe-type schemes by some form of central averaging over 
the Riemann characteristic wave fan (see \cite{NT}, and \cite{KNP} for the
latest developments).
For most applications, even in shock dominated flows, these schemes give 
yet satisfactory results and are surely more economical than Roe-type
schemes, even when third or higher order reconstruction is applied.

In the present paper, two UCT implementations of central-upwind schemes
based on the Harten-Lax-Van Leer HLL \cite{HLL} two-speed flux and 
component-wise reconstruction are considered:

\begin{enumerate}
\item the simplest second order HLL-UCT implementation is performed 
by reconstructing all (potentially discontinuous) variables by
a {\em Monotonized Centered} (MC) Van Leer \cite{VL} limiter. Flux
upwinding is then achieved by a HLL flux formula, as presented below
for the MHD system. Time integration is finally given by the second order
Runge-Kutta scheme. This second order central-type scheme will be here
referred to as MC-HLL-UCT;
\item A third order central scheme, based on the local Lax-Friedrichs (LLF) 
flux and Liu and Osher Convex-ENO (CENO) scheme \cite{LO}, has been already 
presented and tested in our first paper LD. The choice of CENO has been
favored by the following interesting features: 
first, the high order reconstruction algorithm is able to reduce itself 
to minmod-type limiters (MM or MC) at discontinuities, thus strongly 
reducing spurious oscillations typical of component-wise reconstruction
via ENO interpolants; then, a formulation based on point values, rather 
than cell averages, allows the use of purely one-dimensional reconstruction
routines. The same algorithms are used here in a central-upwind scheme 
now equipped with an HLL flux formula, and this new third order central-type 
scheme will be named CENO-HLL-UCT. For the interested reader, we report 
in the Appendix the main computational steps used.
\end{enumerate}

In the following, we define the HLL two-speed formulas appropriate for MHD.
The HLL upwind flux for the fluid components, say $\ff_x$ at $(\xp,y_k,z_m)$,
may be written as
\be
\ff_x^{U}=
\frac{\alpha_x^+\ff_x^E+\alpha_x^-\ff_x^W-\alpha_x^+\alpha_x^-(\uu^W-\uu^E)}
{\alpha_x^++\alpha_x^-},
\label{hll}
\ee
where as usual $\ff_x^a=\ff_x(\ww_x^a,b_x),\,a=E,W$, and
\be
\alpha_x^\pm=\mathrm{max}\{0,\pm\lambda_x^\pm(\ww_x^W,b_x),
\pm\lambda_x^\pm(\ww_x^E,b_x)\},
\label{alpha}
\ee
and similarly for the other $y$ and $z$ components.
Here, to avoid the definition of an intermediate Roe-type state, we have 
chosen to calculate the dissipative $\alpha^\pm_x$ terms by taking the 
maximum eigenvalues (in MHD systems related to the fast magneto-sonic speeds: 
$\lambda_x^\pm=v_x\pm c^f_x$) between left and right states. 

Notice that, by rearranging the terms, it is still possible to rewrite
Eq.~(\ref{hll}) in the form of Sect.~2.3 $\ff^*-\PPhi$, so that the
discussion on $\divn$ of Sect.~2.4 applies unchanged.
On the other hand, we can not use the same composition rules outlined
in Sect.~2.3 to derive Roe-type magnetic fluxes. The correct form of
the $E_z^U$ upwind flux comes out now by averaging over the two overlapping
$x$ and $y$ Riemann wave fans at the $(\xp,\yp)$ edge, thus
\begin{displaymath}
E_z^{U} =\frac{
\alpha_x^+\alpha_y^+E_z^{NE}+
\alpha_x^+\alpha_y^-E_z^{SE}+
\alpha_x^-\alpha_y^+E_z^{NW}+
\alpha_x^-\alpha_y^-E_z^{SW}}{
(\alpha_x^++\alpha_x^-)(\alpha_y^++\alpha_y^-)}
\end{displaymath}
\be
-\frac{\alpha_y^+\alpha_y^-}{\alpha_y^++\alpha_y^-}(b_x^{S}-b_x^{N})
+\frac{\alpha_x^+\alpha_x^-}{\alpha_x^++\alpha_x^-}(b_y^{W}-b_y^{E}),
\label{hll_e}
\ee
where the $\alpha_x^\pm$ and $\alpha_y^\pm$ at edge center points should 
be calculated by taking the maximum characteristic speeds (in absolute
value) among the four reconstructed states, whereas for sake of efficiency
we actually consider the maximum over the two neighboring inter-cell points, 
where these speeds are already at disposal from fluid fluxes computation.

In spite of a different form, Eq.~(\ref{hll_e}) share with Eq.~(\ref{EzROE}) 
all the required upwind properties, that is to be a four-state formula and to 
reduce to the correct 1-D flux for shocks aligned with $x$, $y$, or $x-y$ 
diagonals. The usual settings are instead recovered for LLF fluxes, 
which are the particular cases of the corresponding HLL ones for
$\alpha_x^+=\alpha_x^-=\alpha_x$ and $\alpha_y^+=\alpha_y^-=\alpha_y$,
where now each single $\alpha_i=\mathrm{max}\{\alpha_i^+,\alpha_i^-\}$
speed defines a symmetrical (central) averaged Riemann fan.

It is interesting to notice that the upwind fluxes of Eq.~(\ref{hll}) 
generalize to MHD the formulas of the \cite{KNP} central scheme 
for the Euler system. 
Moreover, Eq.~(\ref{hll_e}) coincides with that defined in the same paper 
for Hamilton-Jacobi scalar equations, which is to be expected since each
component of the induction equation, for a given velocity field and
expressed in terms of the magnetic potential, has exactly the form of
such equations.

\section {Numerical results on test problems}

We present now a standard set of numerical examples to assess accuracy, 
stability and effective divergence-free properties of our UCT schemes 
presented above. In the following tables and plots we shall refer to
these three schemes as POSITIVE-UCT (the flux-limited Roe-type scheme of 
Sect.~3.1), MC-HLL-UCT, and CENO-HLL-UCT (respectively the second and 
third order central-upwind schemes described in Sect.~3.2).
Most of the choices in the following tests have been inspired by
those in \cite{T}, although a direct comparison of UCT with the other
methods for MHD proposed or reported is difficult to achieve, due to the
use of different underlying schemes, and it will be not our main concern here. 
Comparisons will be made instead among our three UCT schemes, and also
with respect to their corresponding non-UCT counterparts, i.e. when 
magnetic field components are discretized and evolved exactly as the 
other fluid variables (the SUP framework), with the same underlying scheme.
We shall refer to these Euler-like versions as {\em base schemes} (BS).
This will enable us to appreciate the effective improvements introduced 
by the use of our UCT method.

In the following sub-sections we shall provide quantitative measures
of the divergence-free properties in our schemes. To distinguish
between the two types of numerical representations discussed in Sect.~2.2.1
and 2.4, here we shall define, for each cell $C\P$, the two quantities
\be
\divn=\sum_iD_i^{(c)}(\sB_i),\quad\quad \divnn=\sum_iD_i(b_i),
\ee
where $D_i^{(c)}$ are the central derivatives defined in Eq.~(\ref{Dc}),
applied to the cell-centered derived data $\sB_i$, and $D_i$ are the
usual two-point divided differences defined in Eq.~(\ref{D}).
As already discussed, while $\divnn$ will be zero to within
machine accuracy for second order UCT schemes, and arbitrarily small
(see the Appendix) for CENO-HLL-UCT (it is simply not defined for the
BS counterparts), the $\divn$ variable may have $O(1)$ jumps at
discontinuities. Recall that in UCT schemes the onset of monopoles is 
actually measured by the $\divnn$ variable, rather than by $\divn$ like in
other CT and non-CT methods for MHD where cell-centered fields are
employed in fluxes.
Notice that, in spite of the use of the staggered magnetic field $\bb$
in the initial conditions and in the computations, the output data 
will be referred to the cell-centered interpolated field $\BB$.

To obtain more accurate results in the various test problems, schemes may
be tuned by choosing each time the most appropriate slope limiter
in the reconstruction routine.
In all simulations, a CFL number of $0.5$ and an adiabatic coefficient
of $\gamma=5/3$ are used, unless differently specified.

\subsection{A convergence test: the oblique CP Alfv\'en wave}

A propagating circularly polarized (CP) Alfv\'en wave is a well known exact 
{\em non-linear} solution of the multidimensional MHD system, and  
therefore it is often used to measure accuracy of a numerical approximation. 
We notice that this test involves only smooth solutions, thus problems 
related to the divergence-free condition are not expected to arise here.
Following \cite{T}, we consider a CP wave propagating on the $(x,y)$ 
Cartesian plane at an angle $\alpha=30^\circ$ relative of the $x$ axis. 
Periodic boundary conditions can be applied if the computational box 
is defined by $0 < x < 1/\cos\alpha,$ and $ 0 < y < 1/\sin\alpha $. 
Let the coordinate along the propagation direction be $\xi=x\cos\alpha + 
y\cos\alpha$ and $\eta=y\cos\alpha- x\sin\alpha$ be the coordinate 
along the transverse direction, then the initial  values of wave variables 
may be given as $v_{\eta}=B_{\eta}=A\sin(2\pi\xi)$ and 
$v_z=B_z=A\cos(2\pi\xi)$, where $A$ measures the wave amplitude. 
The constant parallel components are defined as $v_{\xi}=0$ and 
$B_{\xi}=1$, along with a uniform density $\rho=1$ and pressure $p=0.1$. 
The chosen values correspond to a wave period $T=1$, to a propagation 
Alfv\'enic speed $v_A=1$, and to a sound speed $c_s=\sqrt{\gamma p/\rho}
\simeq 0.4$.

To check numerical accuracy, the evolved solution $\ww(x,y,t)$ at a time 
$t_{max}=nT$, for a given number $n$ of periods, is usually compared to 
the initial condition $\ww(x,y,0)$ and the difference $\delta \ww$ evaluated 
in some norm is then tabulated at different grid resolutions.
To compare with the Toth results, we also adopt the $L_1$ norm
to evaluate the relative errors and the averaged $\bar{\delta}$ values
are measured only by taking into account the transverse vector components
$v_\eta$, $v_z$, $B_\eta$, and $B_z$. However, since we have noticed that
convergence was not precisely achieved with the values $A=0.1$ and $n=5$
suggested by Toth, especially for our most accurate third order central 
scheme, we have decided to use here the safer values $A=0.01$ and $n=1$.
We think that the reason is due to compressible effects: in spite of
the fact that a CP Alfv\'en wave is an exact solution, regardless of
its amplitude $A$, this kind of wave is known to be subject to the
so called {\em parametric decay} instability (see \cite{DVL} and 
references therein). This is due to non-linear wave-wave interactions
that, via coupling to compressive modes, lead to wave distortion and decay,
so that when this happens we are no longer comparing with the true solution.

The results are reported in Table~\ref{table_alfven} for all our three
UCT schemes, with resolutions ranging from $8^2$ to $128^2$. 
In our highest accurate scheme CENO-HLL we use the smoother
MM limiter, while for the second order schemes the sharper MC limiter 
is employed, otherwise results at low resolution are not quite satisfactory,
and a small value of $c_{\rm LLF}=0.01$ is used in POSITIVE.
In cases where only smooth fields are involved, UCT schemes may perform 
worst than their BS counterparts, due to the additional interpolations 
needed to recover the cell-centered magnetic fields (e.g. in output files).
We have verified that in this particular test problem the error for the BS 
versions (not reported here) are approximately $10\%$ less than for their
UCT counterparts.

\begin{figure}
\centerline{\resizebox{.8\hsize}{!}{\includegraphics{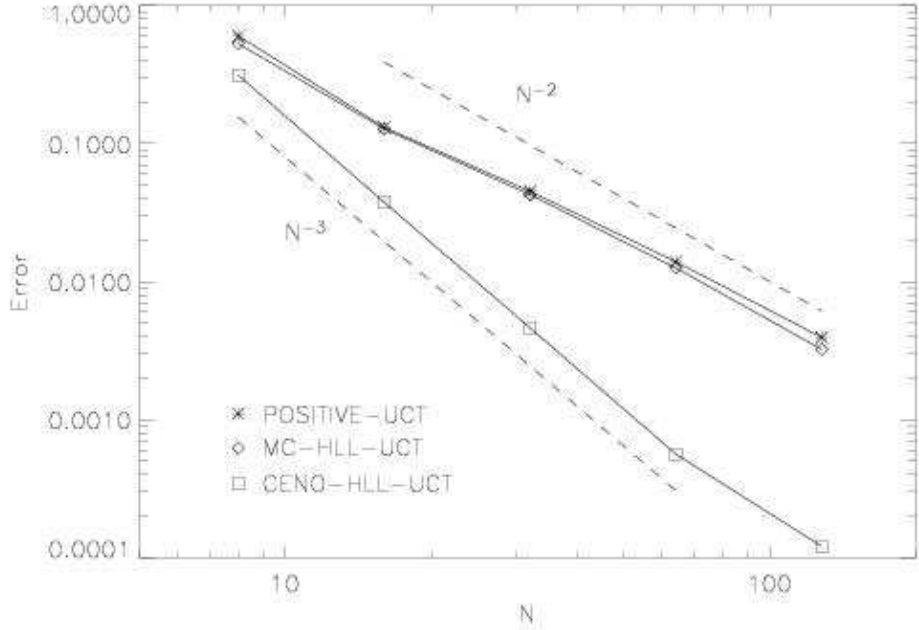}}}
\caption{Convergence test for the oblique 2-D CP Alfv\'en wave problem.
Average $L_1$ errors on transverse $\eta$ and $z$ wave components are shown 
in logarithmic scale for our three UCT schemes. Precise second order 
accuracy is achieved only asymptotically for POSITIVE-UCT and MC-HLL-UCT, due
to the use of the MC limiter which tends to somehow sharpen wave profiles,
while third order convergence is clearly reached by CENO-HLL-UCT, here
employing MM in the reconstruction algorithm, already at very low resolutions.}
\label{plot_alfven}
\end{figure}

The results for all UCT schemes are also plotted in Fig.~\ref{plot_alfven},
where the convergence rates are apparent. Note that in this kind of
problems CENO-HLL is obviously by far the most performing code: the 
accuracy reached with a resolution of $32^2$ is comparable to that 
obtained by the second order methods with $128^2$.

Concerning divergence-free properties, the particular settings of
the problems are such that both $\divn=0$ and $\divnn=0$ to
within machine accuracy for the second order UCT schemes, because
the invariance direction $\eta$ is made to coincide with the cell
diagonal, a condition required by the double periodicity (independently
on the value of the angle $\alpha$). For CENO-HLL-UCT, typical values
for the maximum values of $\divn$ and $\divnn$ are $10^{-8}$ and
$10^{-12}$, respectively.

\subsection{Rotated shock-tube problems}

These test problems involve the propagation of discontinuities defined by
usual 1-D shock-tubes on a 2-D computational plane, and are relevant to 
some main aspect considered here on the divergence-free condition.
However, specific divergence-free properties may be hidden if one
(or both) of the following special conditions hold:
\begin{enumerate}
\item the initial magnetic field is uniform;
\item the propagation direction lies along cell sides or diagonals.
\end{enumerate}

In the first case it is clear that any representation of $\divn$
will be exactly zero for initial fields. Then its subsequent time
evolution will only give a measure of the ability of a numerical 
scheme to preserve the initial $\divn$, even if it is non-vanishing,
while a characterizing aspect of any CT-based scheme is precisely the 
possibility to deal with discontinuous divergence-free fields.

Concerning the second case, the problem is the same as in the previous
test, though here involves discontinuities and will be described in details.
For initial symmetric conditions where all variables are defined 
as $w(x,y)=w(\xi)$, where $\xi$ is a coordinate making an angle $\alpha$ 
with respect to the $x$ axis, and they do not depend on the transverse 
$\eta$ coordinate, it is important to check at later times not only the
evolution properties along the $\xi$ coordinate but also the conservation 
of the transverse invariance. In particular, the $\divb=0$ condition expressed 
in the rotated coordinates is given by
\be
\divb = \partial_{\xi}B_{\xi}+\partial_{\eta}B_{\eta}=0,
\ee
and the equivalent condition $B_{\xi}(x,y)=\mathrm{const}$ can be recovered 
only if $B_{\eta}$ (and all other variables, of course) are $\eta$ independent.
However, in numerical applications based on standard Cartesian grids this
condition may be achieved only if the $\xi$ and $\eta$ directions are
aligned with the cell diagonals (or the cell sides). In fact, any 
discontinuity front making a different angle will be discretized with unequal 
jumps along $x$ and $y$. Thus, even when the $\divnn=0$ condition holds,
the $B_{\xi}=const$ relation does so only in approximate sense, with $O(1)$ 
jumps where discontinuities occur.
More strictly related to the errors in the $B_\xi$ variable, which has
to be necessarily calculated from the interpolated $\sB_i$ cell-centered 
fields, is the $\divn$ variable, which in fact shows the highest jumps
precisely at discontinuities. We have verified that, when the angle
$\alpha=45^\circ$ is chosen, both the conditions $B_\xi=const$ and $\divn=0$
hold within machine accuracy, as in the previous test.

The numerical domain for the oblique shock tube tests may be reduced, as 
cleverly suggested by Toth, to just a narrow strip $[0,1]\times [0,2/N]$, 
discretized with a $N\times 2$ grid (so that $dx=dy$). {\em Shifted} 
boundary conditions in the $\eta$ direction are applied and 
$\alpha=\tan^{-1} 2\approx 63.4^\circ$. Each 2-D run, performed with $N=256$, 
is compared with the corresponding 1-D test on a $1024$ grid,
by using the data along the $x$ axis (the first row) at a final time 
$t_{\rm max}\cos\alpha$ (where $t_{\rm max}$ refers to the 1-D test).

The initial left (L) and right (R) states of the three shock tube
problems considered here are reported in Table~(\ref{table_st_lr}),
and the final times are $t_{\rm max}=0.08$ for ST-1,  $t_{\rm max}=0.2$ for 
ST-2, and $t_{\rm max}=0.1$ for ST-3. ST-1 is a coplanar 2-D problem 
of converging shocks in an initially uniform magnetized background, 
ST-2 is a non-coplanar case involving Alfv\'enic discontinuities,
and ST-3 is the famous (coplanar) Riemann problem involving the so-called
compound (or intermediate) shock. Note that in ST-2 and ST-3 the magnetic
field has jumps in the initial data and $\alpha\neq 45^\circ$, so none
of the special cases indicated above apply. Other Riemann problems have been
checked and the UCT schemes appear to behave well in all cases, including for
example non-coplanar tests with compound shocks. The same three problems have 
been tested in LD for the CENO-LLF-UCT code on a symmetric $N\times N$ grid
with $\alpha=45^\circ$, while Toth shows just the first two tests: ST-1 
with precisely the same settings and ST-2, the non-coplanar 2.5-D problem, 
with $\alpha=45^\circ$. 

\begin{figure}
\centerline{\resizebox{15cm}{7cm}{\includegraphics{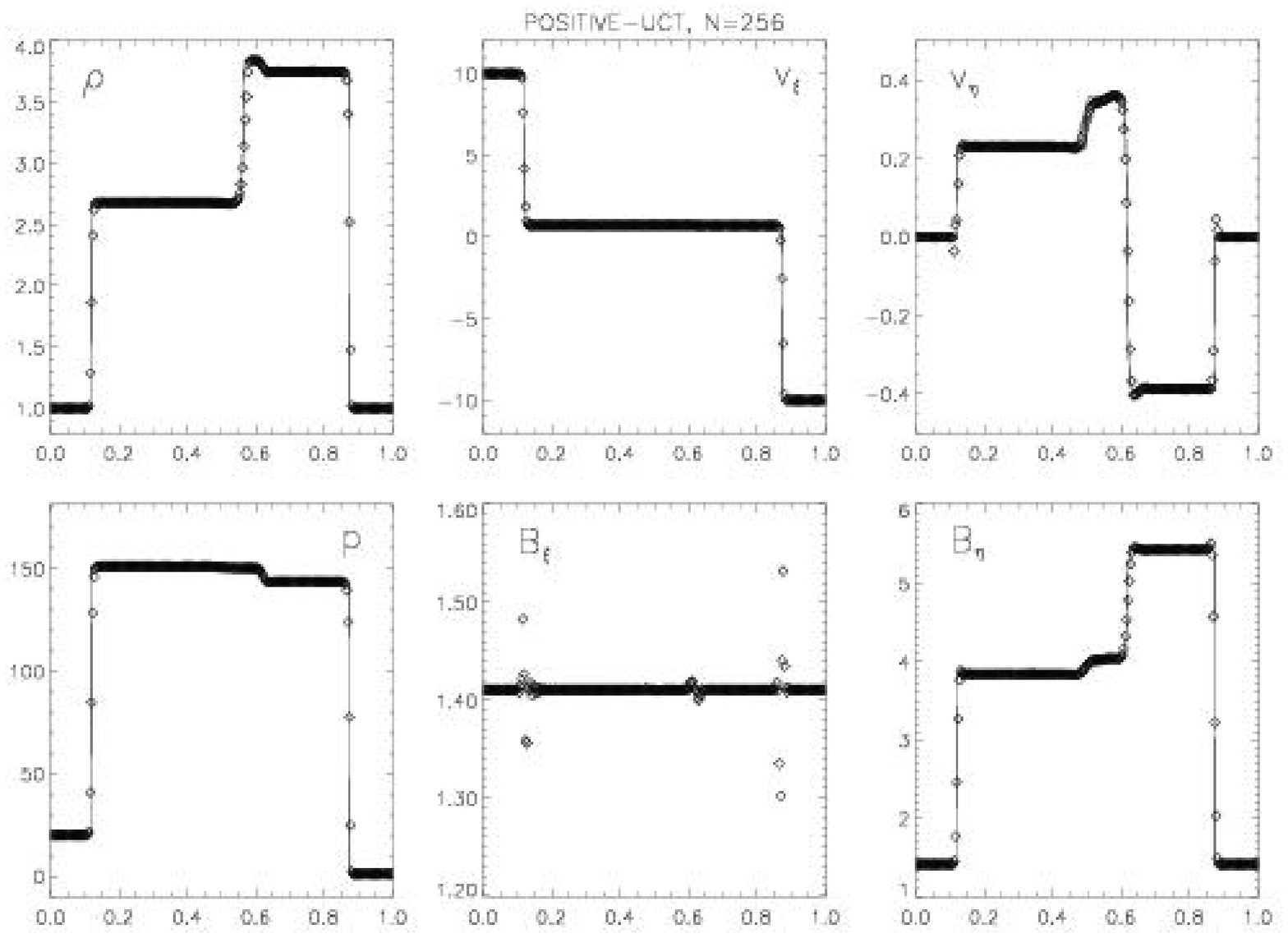}}}
\vspace*{5mm}
\centerline{\resizebox{15cm}{7cm}{\includegraphics{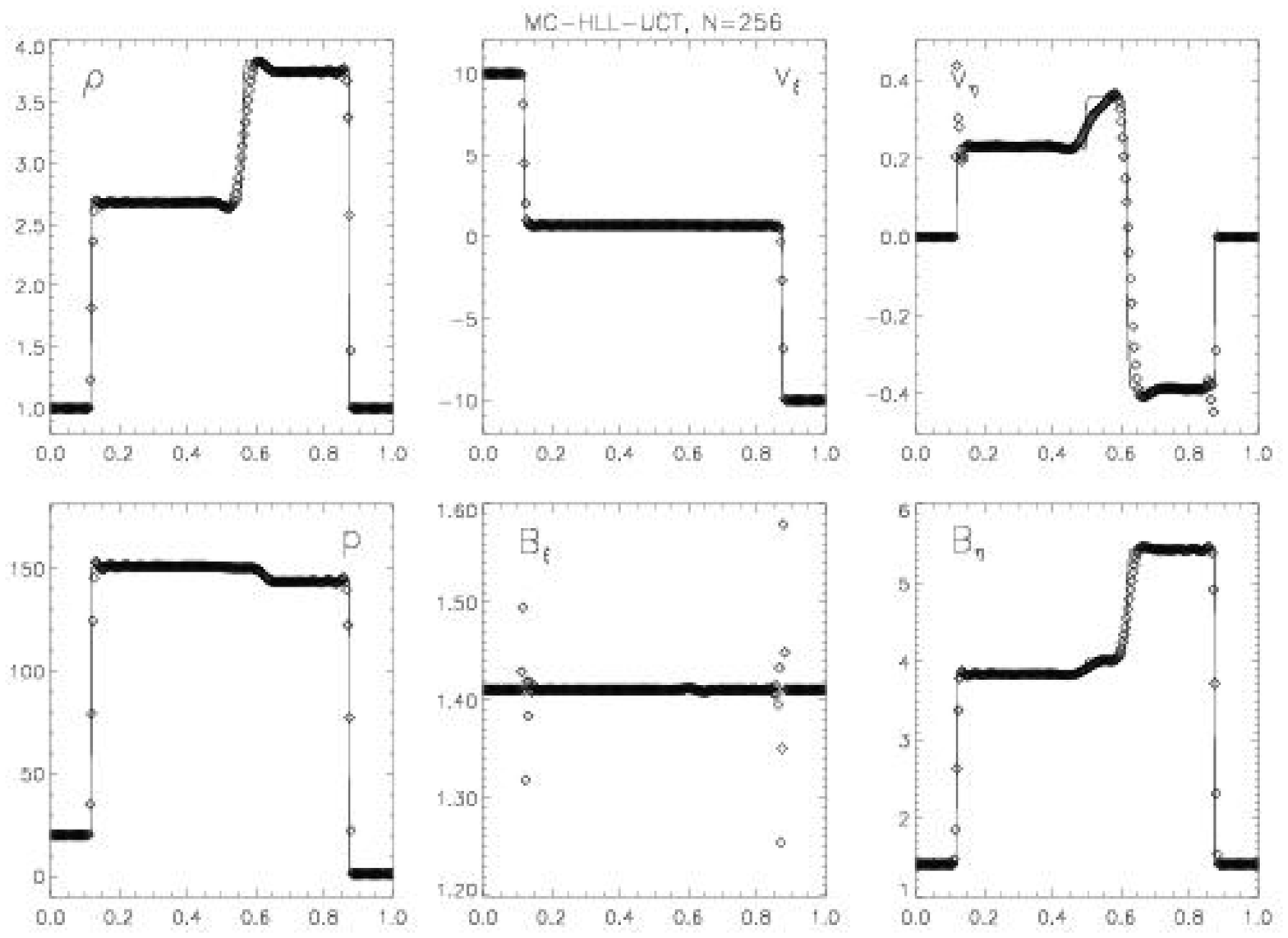}}}
\vspace*{5mm}
\centerline{\resizebox{15cm}{7cm}{\includegraphics{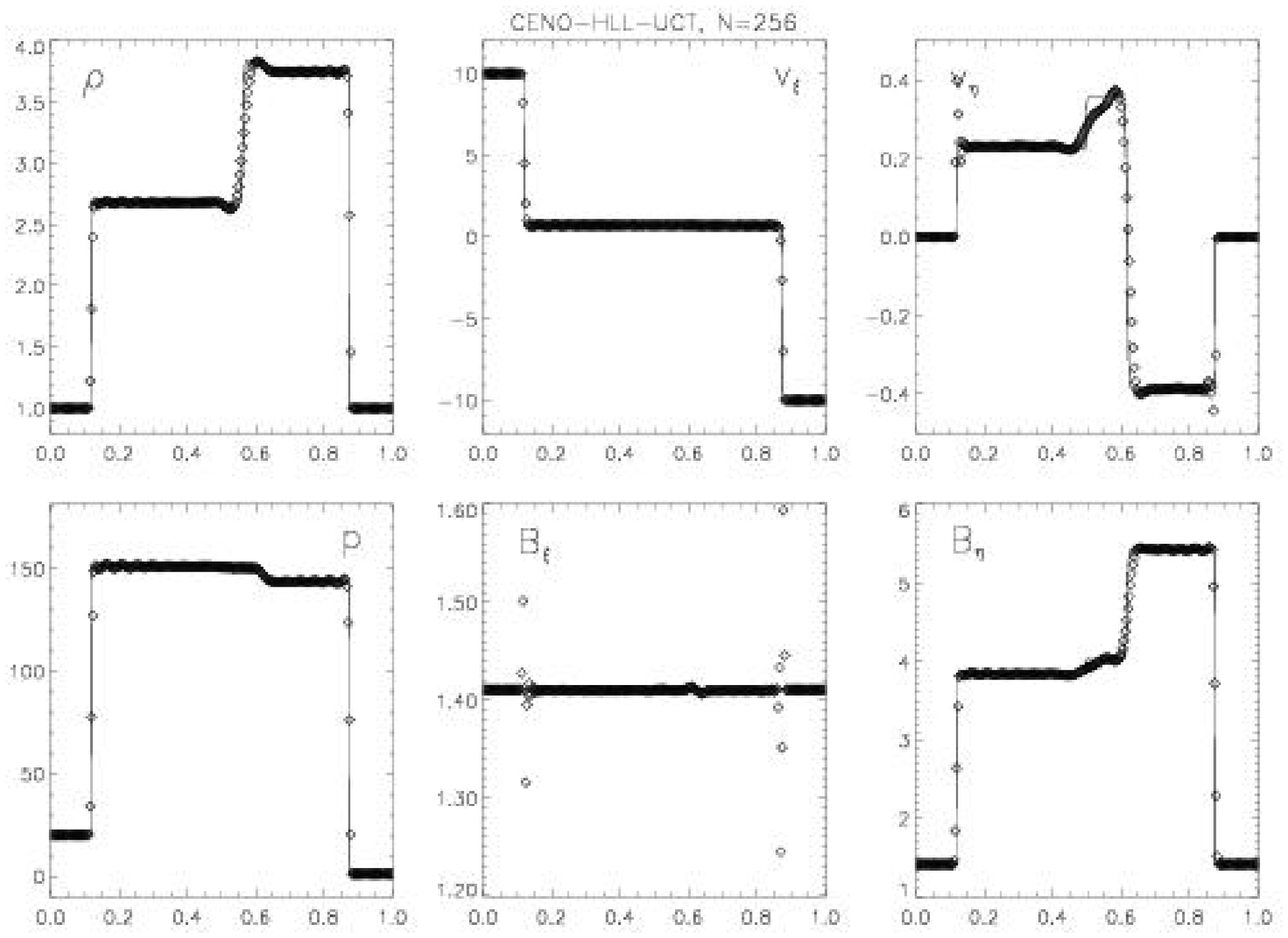}}}
\caption{The oblique ST-1 shock tube problem. The numerical results from
our three UCT schemes obtained on a $256\times 2$ grid (symbols) are 
compared with the 1-D solution on a $1024$ grid (solid line).}
\label{plot_st1}
\end{figure}

\begin{figure}
\centerline{\resizebox{15cm}{7cm}{\includegraphics{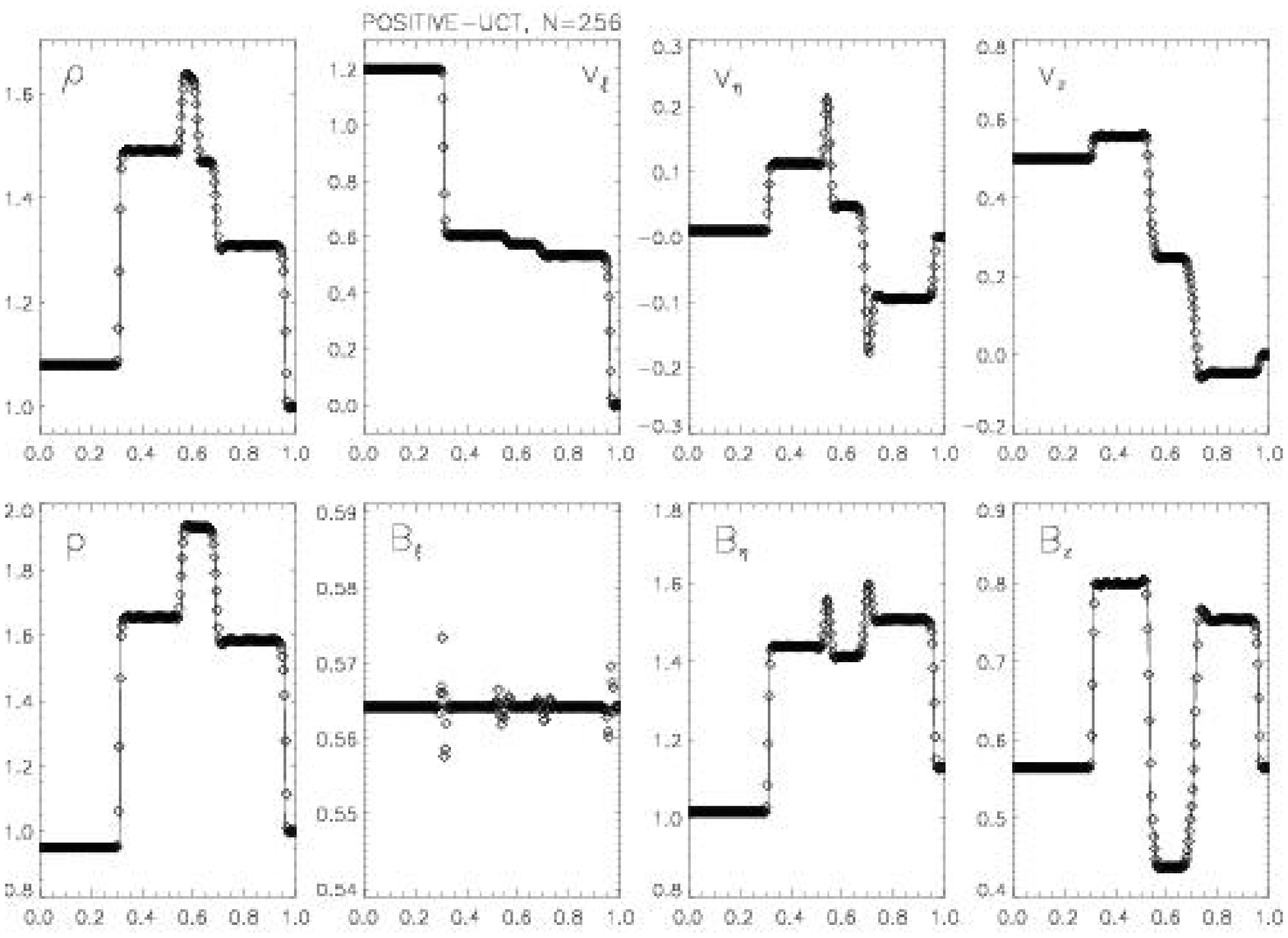}}}
\vspace*{5mm}
\centerline{\resizebox{15cm}{7cm}{\includegraphics{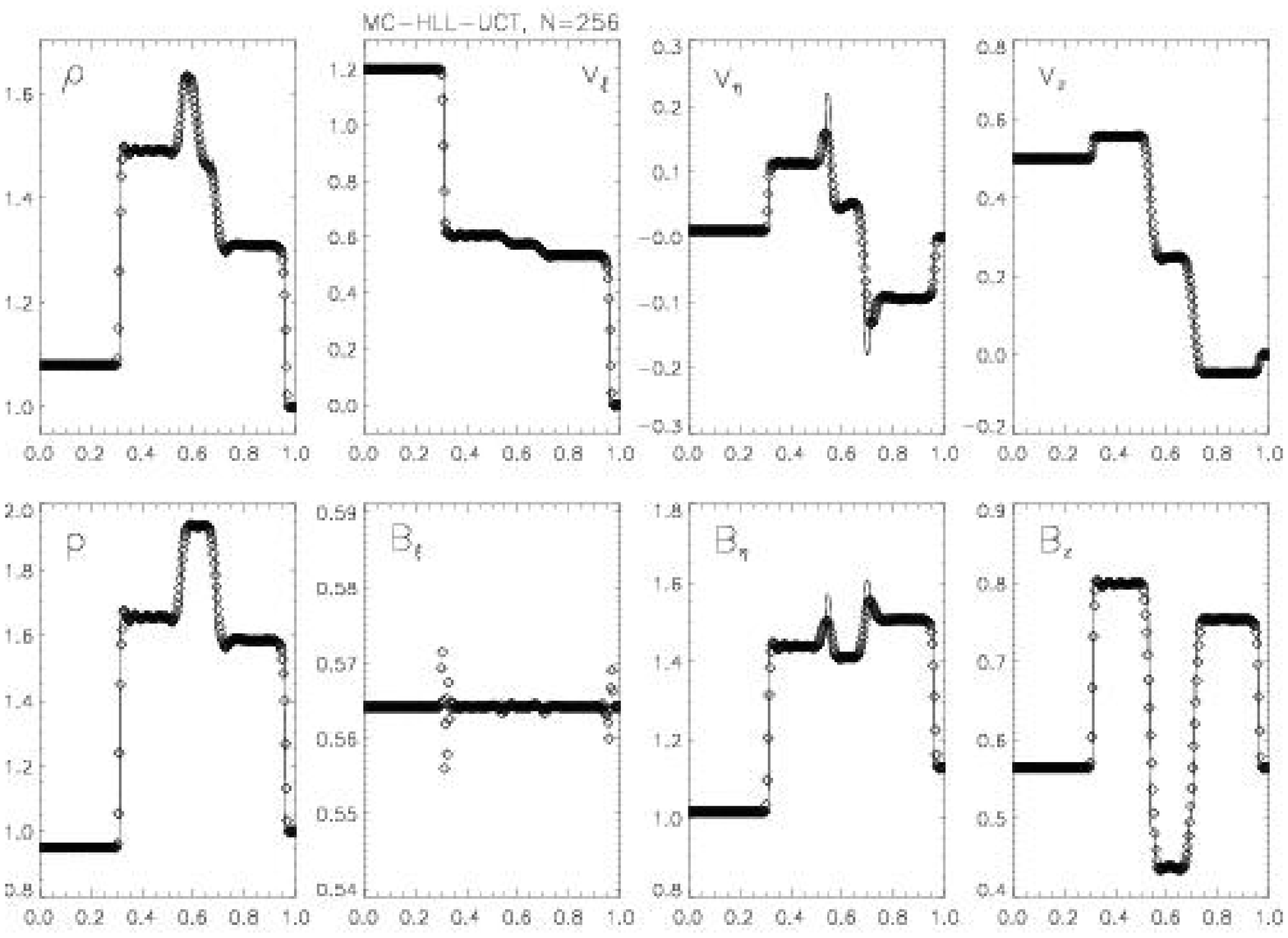}}}
\vspace*{5mm}
\centerline{\resizebox{15cm}{7cm}{\includegraphics{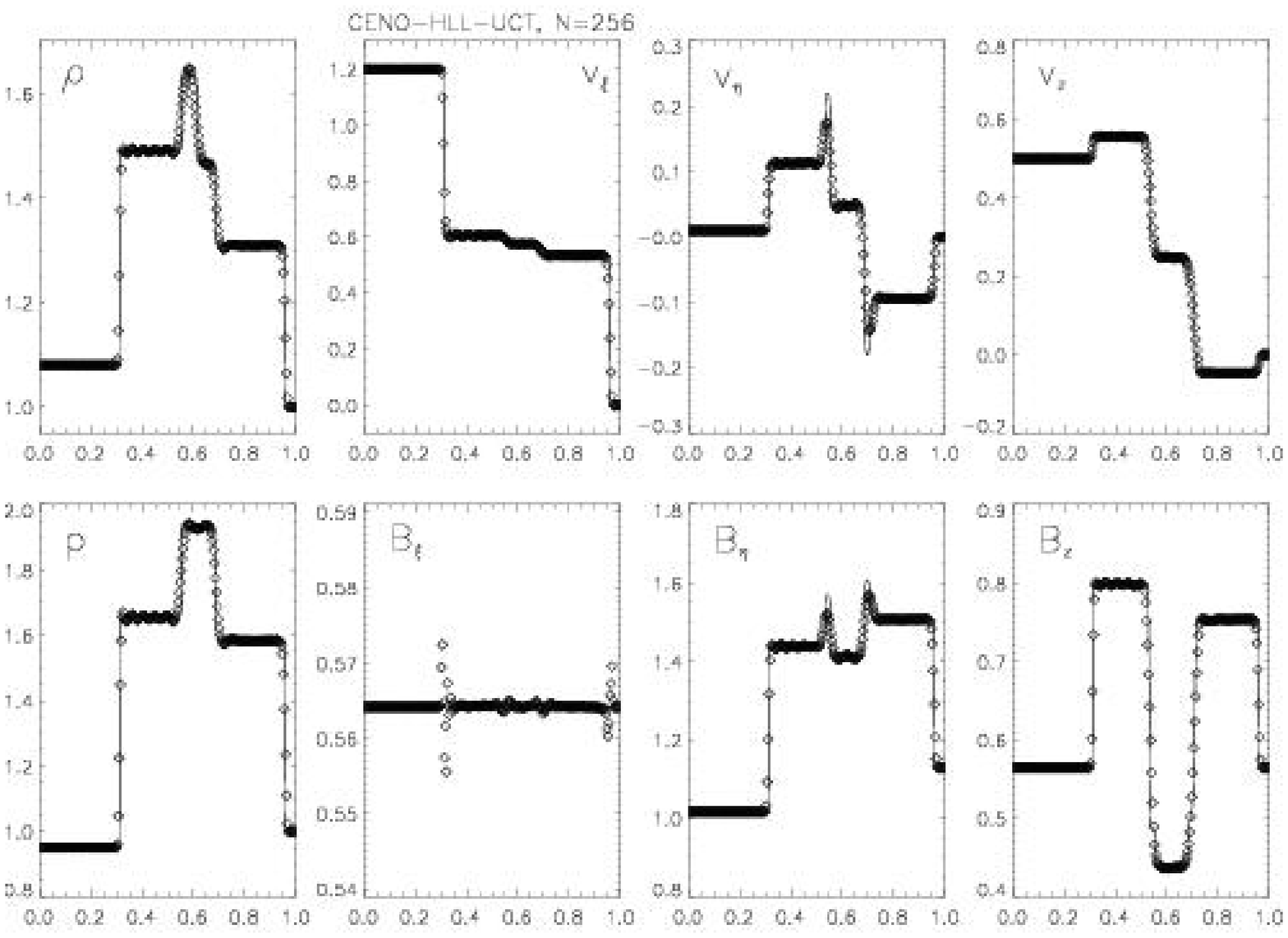}}}
\caption{The oblique ST-2 shock tube problem. The numerical results from
our three UCT schemes obtained on a $256\times 2$ grid (symbols) are 
compared with the 1-D solution on a $1024$ grid (solid line).}
\label{plot_st2}
\end{figure}

\begin{figure}
\centerline{\resizebox{15cm}{7cm}{\includegraphics{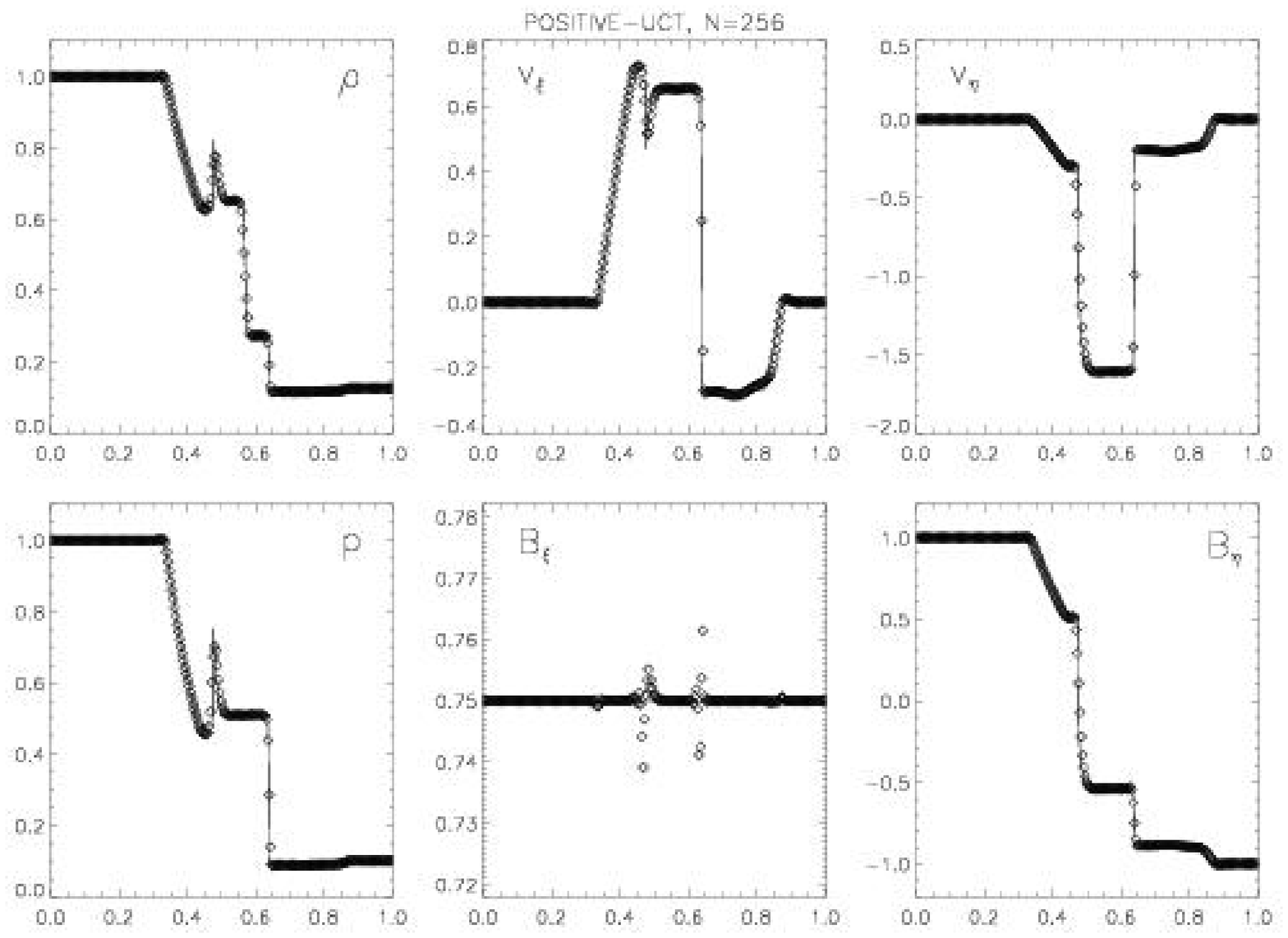}}}
\vspace*{5mm}
\centerline{\resizebox{15cm}{7cm}{\includegraphics{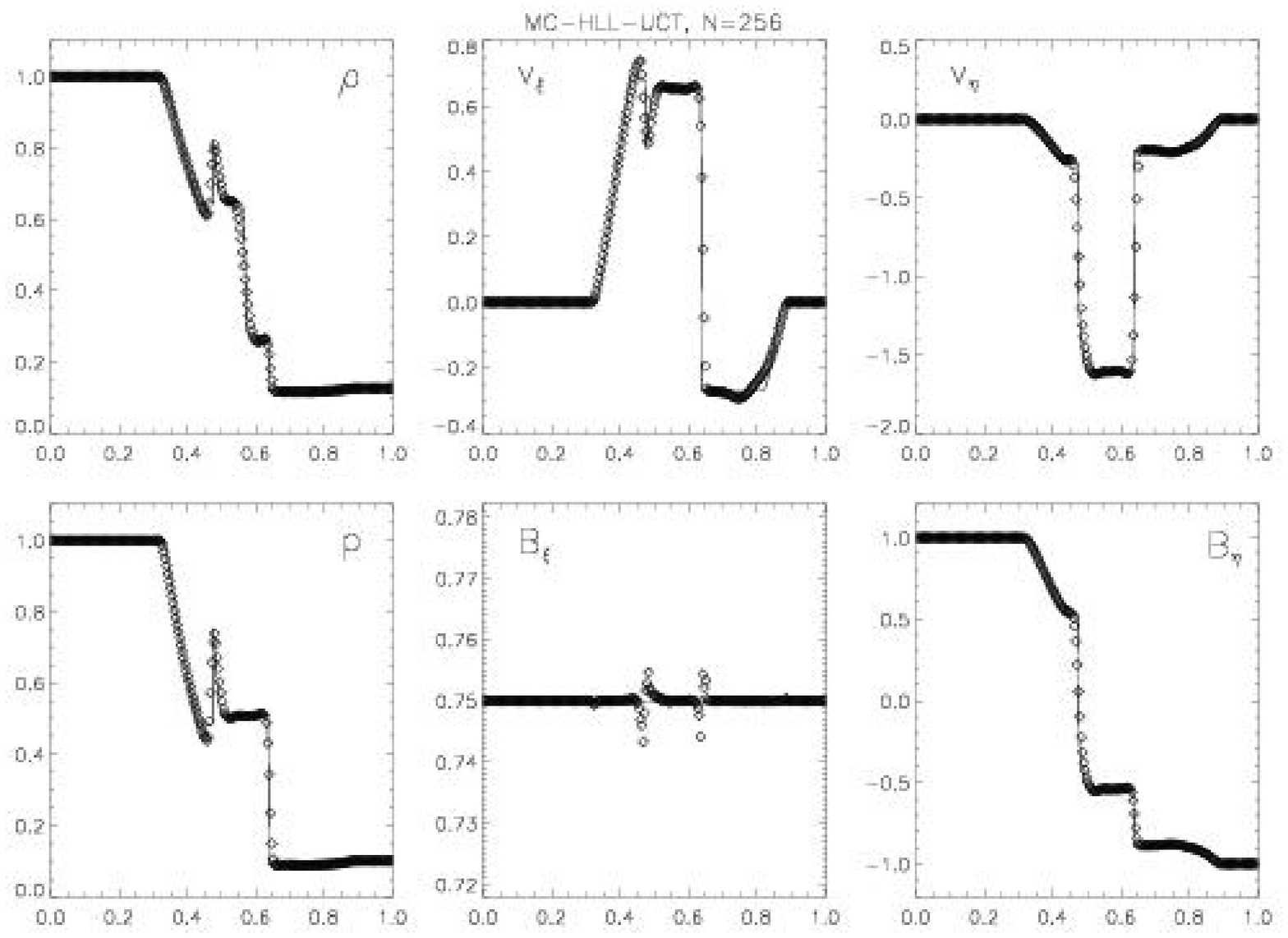}}}
\vspace*{5mm}
\centerline{\resizebox{15cm}{7cm}{\includegraphics{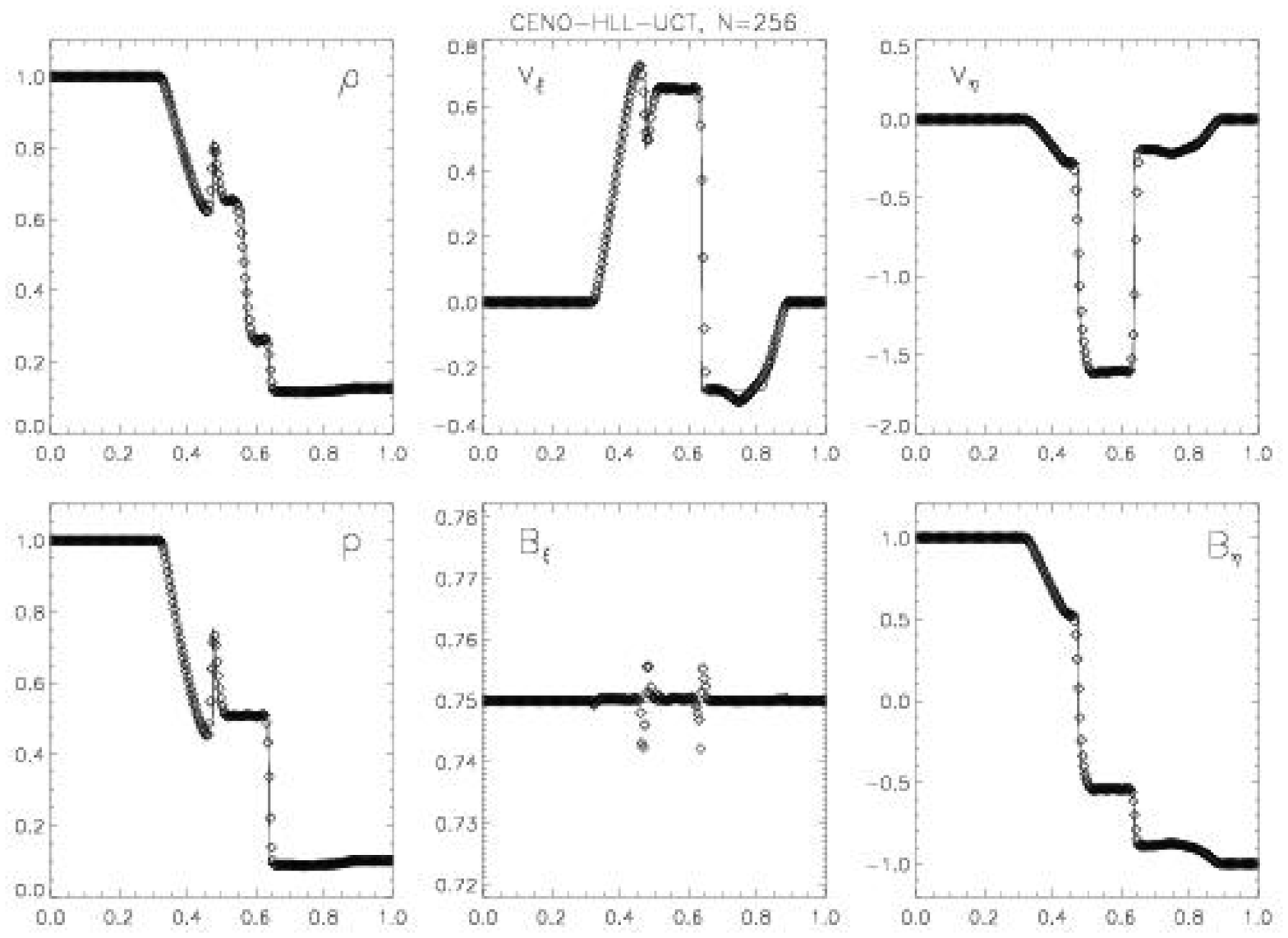}}}
\caption{The oblique ST-3 shock tube problem. The numerical results from
our three UCT schemes obtained on a $256\times 2$ grid (symbols) are 
compared with the 1-D solution on a $1024$ grid (solid line).}
\label{plot_st3}
\end{figure}

Here the most steepening slope limiters and a minimum amount of viscosity
are used in our schemes: thus POSITIVE uses the {\em Superbee} (SB) limiter 
for all entries of $L$, $c_{\rm LLF}=0.01$, and the limiter in CENO-HLL is MC.
In Table~(\ref{table_st}) we report, for all tests and numerical schemes, 
the $\ad$ average L1 norm over the variables involved in each test,
the L1 norm of the errors in $B_\xi$, the L1 norms of variables
$\divn$ and $\divnn$ (respectively $[\divb]_{\rm avr}$ and 
$[\divbb]_{\rm avr}$), and their maximum absolute value over the 
computational domain (respectively $[\divb]_{\rm max}$ and 
$[\divbb]_{\rm max}$).
The comparison between the $x$ projection of the evolved quantities
and the reference 1-D runs for our three UCT schemes are plotted in 
Fig.~(\ref{plot_st1}), Fig.~(\ref{plot_st2}), and Fig.~(\ref{plot_st3}), 
for the ST-1, ST-2, and ST-3 tests, respectively.  

For problems involving sharp discontinuities, like shock tubes,
schemes based on characteristics are clearly preferable since
sharp limiters, which lead to more accurate results, may be often 
used there without producing spurious oscillations. This can be appreciated
from the plots and may be measured more quantitatively from the 
reported errors in Table~(\ref{table_st}), which are the lowest for our
POSITIVE scheme. If limiters less sharp than SB are used in POSITIVE-UCT
the errors obviously increase: with MC we find $\ad=0.0146$ for ST-1,
$\ad=0.0153$ for ST-2, and $\ad=0.0203$ for ST-3; with the smoothest
MM we find $\ad=0.0201$ for ST-1, $\ad=0.0225$ for ST-2, and
$\ad=0.0298$ for ST-3. Concerning the central schemes, the use of
the MC limiter in the reconstruction step allows to produce accurate 
results with a rather low level of oscillations even in the absence of
characteristics decomposition. These schemes are just less accurate
at contact and Alfv\'enic discontinuities, since the related characteristic
waves do not enter the HLL flux definition.

\begin{figure}
\centerline{\resizebox{15cm}{7cm}{\includegraphics{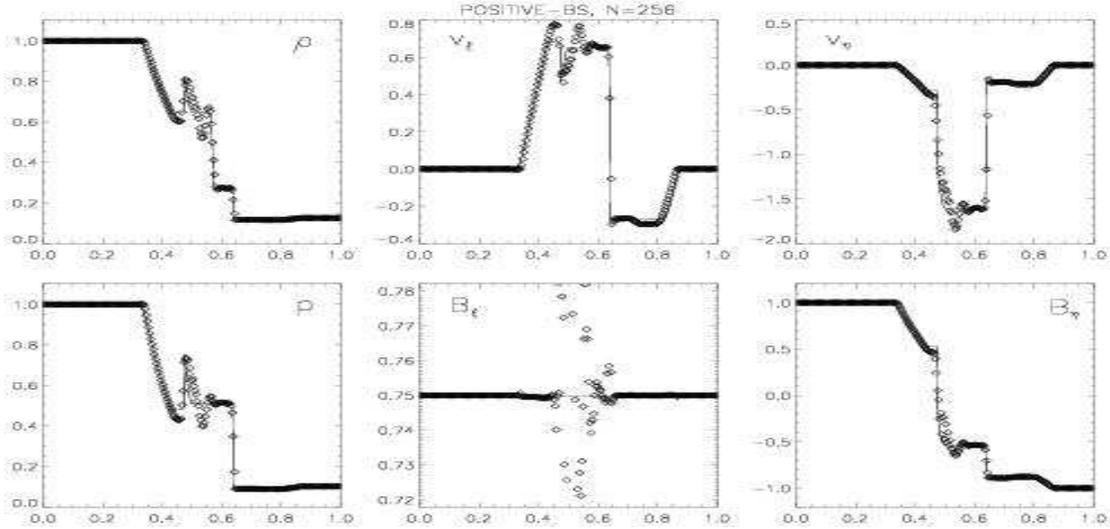}}}
\caption{The oblique ST-3 shock tube problem, this time for the base
scheme (BS) version of POSITIVE. Note the presence of errors induced 
by the numerical monopoles, here free to arise and grow in time.}
\label{plot_st3_bs}
\end{figure}

Finally, to appreciate the level of damage that monopoles can produce,
we plot in Fig.~(\ref{plot_st3_bs}) the results of POSITIVE-BS for ST-3,
where monopoles are free to arise (even in initial data since here none
of the two special conditions apply) and grow. By also looking at the tabulated
errors, it is clear that BS results are systematically worst with respect to 
the correspondent UCT versions (see in particular the errors on $B_\xi$).

\subsection{The Orszag-Tang vortex problem}

A well-known model problem to study the transition to MHD turbulence
is provided by the so-called Orszag-Tang vortex, which has been later
adopted as a standard 2-D test for MHD shock-capturing codes.
The initial conditions are here $v_x=-\sin y$, $v_y=\sin x$,
$B_x=-\sin y$, $B_y=\sin 2x$, $\rho=\gamma^2$, $p=\gamma$ (so the the
sound speed and the initial Mach number are both 1), $v_z=B_z=0$.
The computational domain is a square $0<x,y<2\pi$ $N\times N$ box with 
periodic boundary conditions along both directions, while the output time
is $t_{max}=\pi$ (note that in LD the magnetic field was normalized
against $B_0=1/\sqrt{4\pi}$).

\begin{figure}
\centerline{\resizebox{15cm}{22cm}{\includegraphics{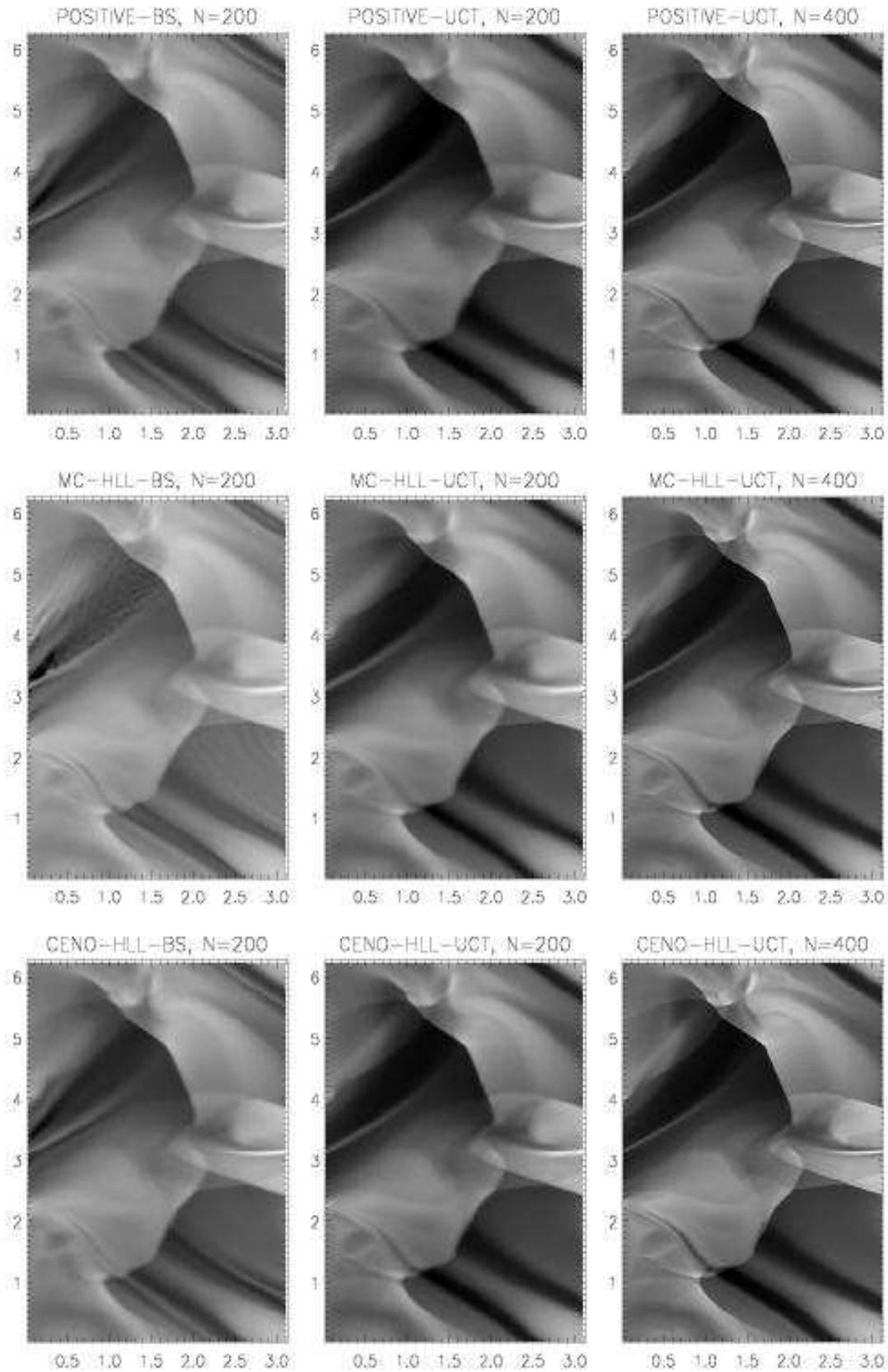}}}
\caption{Gray-scale images of the temperature $T=p/\rho$ distribution
in the Orszag-Tang vortex problem. For each scheme, low resolution
($200^2$) results for the BS and UCT versions are compared with
the corresponding high resolution ($400^2$) UCT reference run.}
\label{plot_ot}
\end{figure}

All our UCT schemes have been tested with the MC limiter ($c_{\rm LLF}=0.1$
in POSITIVE), at the resolutions of $50^2$, $100^2$, $200^2$, and $400^2$.
In Table~(\ref{table_ot}) we report the averaged errors (over the six
evolving variables). For each scheme, errors are measured with respect to a 
highest accuracy run at $400^2$, obtained as the average of the results
from the three UCT schemes.
More qualitative comparisons can be appreciated in Fig.~(\ref{plot_ot}),
where gray-scale images of the temperature for the $200^2$ runs, for
both the BS and UCT versions, are compared to the respective reference 
solution (the UCT run with $400^2$ grid points). Due to the symmetry of
the problem, only the $[0,\pi]\times [0,2\pi]$ first half of the domain
is displayed.
The first thing to be noticed is that the BS versions clearly produce
incorrect results: the darkest, rather homogeneous features of the
reference solutions appear more structured, with whiter filaments, and
sometimes spurious oscillations are visible.
By also reading Table~(\ref{table_ot}), it is then quite apparent that
the three schemes behave similarly in this context. Due to the numerous
steepened structures the order of accuracy is close to one in all cases,
so that the errors are very similar. CENO-HLL-UCT is the most performing
scheme, giving results almost identical to the Roe-type POSITIVE-UCT, while
MC-HLL-UCT is of course more dissipative, but its results are still
satisfactory. 

\subsection{The fast rotor problem}

In \cite{BS} a model problem to study the onset and propagation of
strong torsional Alfv\'en waves was presented and analyzed.
A disk of radius $r=0.1$ made up of dense fluid ($rho=10$) rotates with high
angular velocity ($\omega=20$) in a static, magnetized ($B_x=5/\sqrt{4\pi}$)
background with uniform density and pressure ($\rho=p=1$). The adiabatic index
is $\gamma=1.4$. To conform with \cite{T}, the final time is $t=0.15$ and the
same initial {\em taper} function is used (note that in LD the same problem
was solved by the CENO-LLF-UCT code with $t=0.18$ and no tapering).

\begin{figure}
\centerline{\resizebox{16cm}{16cm}{\includegraphics{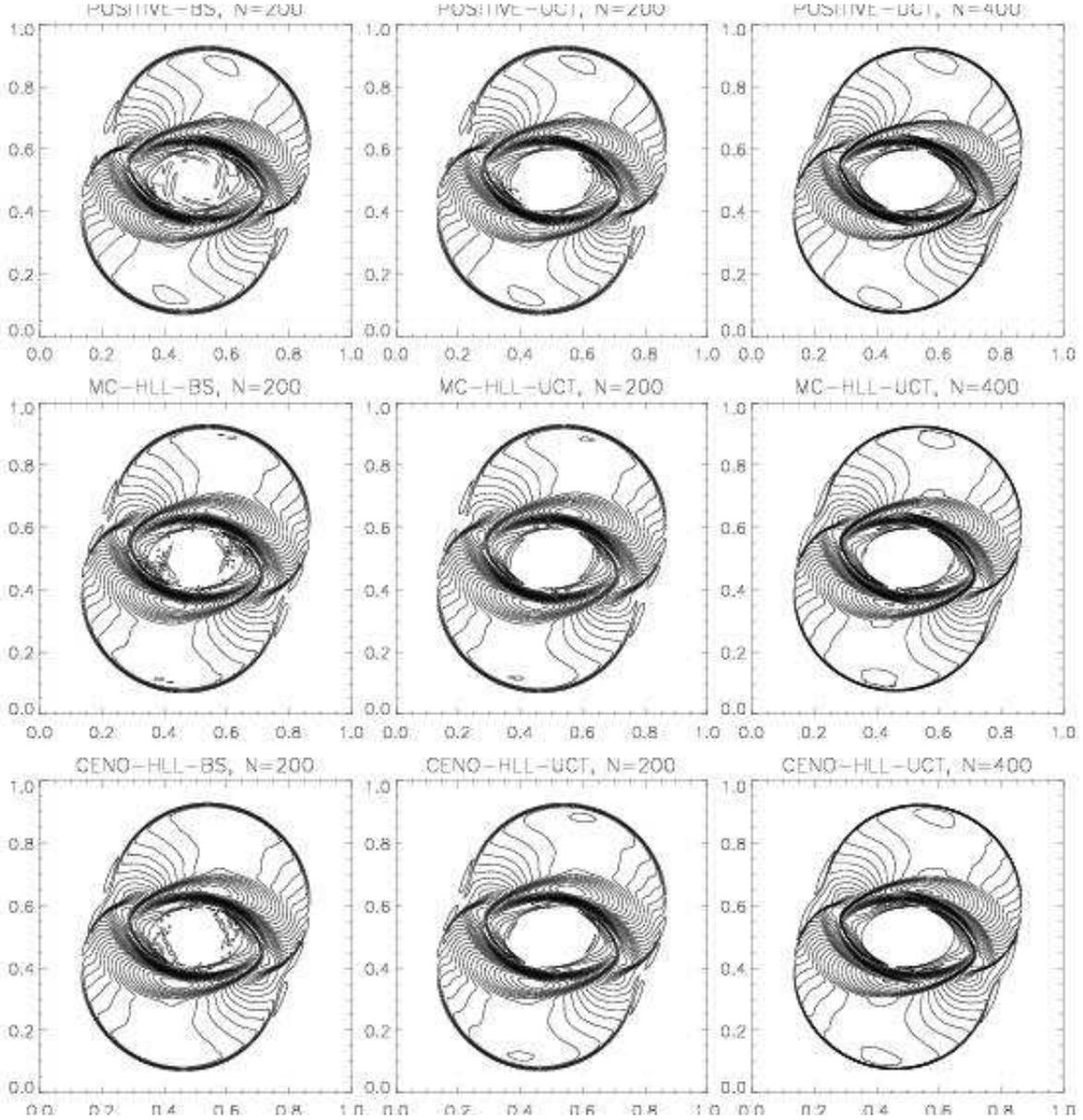}}}
\caption{Contours of the magnetic pressure $p_m=B^2/2$ distribution
in the fast rotor problem. For each scheme, low resolution
($200^2$) results for the BS and UCT versions are compared with
the corresponding high resolution ($400^2$) UCT reference run.}
\label{plot_rotor}
\end{figure}

The numerical settings are identical to those employed in the previous test,
and also the errors displayed in Table~(\ref{table_rotor}) are calculated
in the same way. In Fig.~(\ref{plot_rotor}) the magnetic pressure
$p_m=B^2/2$ is shown as isocontours diagrams for all schemes, first
the BS and UCT runs at $200^2$ grid points, compared with the corresponding
UCT reference run at $400^2$, which is also used to calculate the errors
for the lower accuracy tests.
Here the same remarks made above still apply: the accuracy order is low
due to discontinuities, the two second order schemes behave
very similarly (thus MC-HLL-UCT has to be preferred for its efficiency),
while CENO-HLL-UCT gives the sharpest profiles, thus in this kind of model
problems the accuracy in the reconstruction seems to be more important
than the accuracy in resolving the Riemann structures.
To conclude, we also notice that the BS schemes appear to behave
correctly far from the rotating disk, where the waves are propagating
outwards, while a lot of numerical noise is clearly formed inside the
disk, where the numerical monopoles have time to accumulate, probably
in a way similar to the inclined shock tube problems discussed in Sect.~4.2.

\section{Conclusions}

We have presented a method, first outlined in \cite{LD},
to construct Godunov-type schemes for the MHD system, 
named {\em Upwind Constrained Transport} (UCT). 
The main intent of our work is to assure that specific properties
of the magnetic field, related to the basic divergence-free relation, 
enter as a {\em built-in} properties also in the approximated systems.
To that purpose, by taking advantage of the CT discretization technique, 
we have presented specific procedures to define consistent
derivative approximations, reconstruction steps and approximate
Riemann solvers {\em all based} on the staggered (or face centered) magnetic
field components $b_i$ chosen as primary data. A main advantage on this
approach is that no cleaning procedures or {\em ad hoc} modifications
of the form of the MHD conservative system are required.

The main computational steps entering a UCT-based schemes are:
\begin{enumerate}
\item reconstruction procedures based on the smoothness properties
of the divergence-free $\BB$ vector field, as represented in
finite volume CT discretization (Sect.~2.2);
\item the application of standard approximate Riemann solvers for
the momentum and energy equations, with the prescription that only 
variables not related to the divergence-free condition are reconstructed 
and participate to the upwind differentiation. As a benefit, among others,
exact cancellation of numerical monopoles is assured (Sect.~2.3 and 2.4);
\item a specific formulation of the approximate Riemann solvers for
the induction equation (Sect.~2.3);
\item a time integration procedure where no time-splitting is adopted.
\end{enumerate}

To demonstrate the validity and flexibility of our UCT method, we have
finally applied it to a flux-limited Roe-type scheme (the {\em positive}
scheme by Liu and Lax \cite{LL1}), which proves to be accurate, robust
and well-suited for more demanding applications requiring AMR techniques.
This novel scheme has been then tested numerically on a standard set
of model problems and compared to central-type second and third order 
schemes based on the two-speed HLL solver.

We conclude by remarking that our method, defined here for the classical
MHD system in regular structured grids, applies unchanged to the equations of 
special and general relativistic MHD (see \cite{DBL}), and many procedures 
here presented may have a natural generalization for grid refinements and 
unstructured grids.

\vskip 5mm
\noindent
{\em Acknowledgments} The authors thank G. Toth and another reviewer for
their competent help in improving the manuscript.
 
\appendix
\section{Point-value formalism and third order procedures in the
CENO-HLL-UCT scheme}

As shown in Sect.~2.3, to get high $r\geq 3$ order schemes
in a finite-volume setting, besides a
proper $r-$th order reconstruction of $\aww$ variables, a final
averaging of the $\aff_i$ flux is needed. In the 3-D case the latter
procedure is not cost-effective and more efficient implementations
have been proposed for ENO-type schemes by Shu and Osher \cite{SO}.
In this approach, point values $\uu$, instead of cell averages $\auu$,
are advanced in time and flux values at the interfaces are directly
reconstructed at the desired order by using flux point values $\ff_i(\uu)$.
However, when applied to the MHD system (see \cite{JW}), this implementation
is not suitable to take properly into account the divergence-free condition,
since the flux reconstruction yet involves the $B_i$ cell-centered values.
Therefore, as in our previous work (LD), we use more appropriate flux
reconstruction techniques. Let then consider the MHD equations, now in the
form that comes out by applying the inverse operation of volume
averaging to Eqs.~(\ref{3a}) and (\ref{3b}):
\be
\frac{{\rm d}}{{\rm d}t}[\uu(t)]+\sum_i D_i(\hff_i)=0,~~~~
\label{4a}
\ee
\be
\frac{{\rm d}}{{\rm d}t}[\hb_i(t)]+
\sum_{j,k}\epsilon_{i,j,k} D_j(\hE_k)=0.
\label{4b}
\ee
Here $\uu=\uu_{j,k,m}$, now cell-centered point values, constitute the new
set of fluid primary data, and correspondingly the conservative flux
two-point differences $D_i(\hff_i)$ are high order approximations of
point-value first derivatives. In the induction equation,
on the other hand, $\hb_i$ do {\em not} coincide with point-value
representations of the staggered magnetic field components, those
named $b_i$ which are needed in flux computations, since their volume
averages must now return the {\em surface}-averaged $\ab_i$ values.
Therefore, $\hb_i$ components are actually defined, now as primary
data, in the same way $\hff_i$ fluxes are defined in Eq.~(\ref{4a}),
thus two-point differences $D_i(\hb_i)$ give high order representations
of point-value parallel first derivatives. Similarly, $D_j(\hE_k)$
will be here high order representations of the staggered electric fields
first derivatives, and their volume average must give back line-averaged
electric fields.

It is important to notice that if $\hb_i$ components are evolved as
primary data from Eq.~(\ref{4d}), the point-value version of the
divergence-free relation (\ref{divb}), which is written as
\be
\divnn=D_x(\hb_x)+D_y(\hb_y)+D_z(\hb_z)=0,
\label{div2}
\ee
will be preserved in time (if valid at $t=0$) to within machine accuracy,
exactly as in the finite-volume framework of Sect.~2.1.

Our CENO-HLL-UCT scheme is then based on the following steps:

\begin{enumerate}

\item For given $\ww_i$ cell-centered point values we reconstruct
the face left-right point values using the appropriate CENO algorithm.
The upwind fluxes $\ff^U_i(\ww_i,b_i)$ and $E_k^U(\ww)$ are then evaluated
by using HLL as in Sect.~2.3.

\item  For given values of these $\ff_i$ point-value representation of fluxes,
the $\hff_i$ data are defined, for $i=x,y,z$ and at the same points, as
\be
\hff_i=\ff_i-\frac{1}{24}\sD_i^{(2)}(\ff_i)
\label{D2}
\ee
where $\sD_i^{(2)}$ is a non-oscillatory approximation of the
second derivative in the indicated coordinate. In this way, the
difference $D_i(\hff_i)$ provides a high order $r\ge 3$ accurate
approximation of the flux first derivative.
Concerning the magnetic fluxes, given the point values $E_k$ at edge
centers, the corresponding $\hE_k$ data must be defined as
\be
\hE_z=E_z-\frac{1}{24}[\Dx^{(2)}(E_z)+\Dy^{(2)}(E_z)],
\label{hE_z}
\ee
and similarly for $x$ and $y$ components.

\item
Eqs.~(\ref{4a}) and (\ref{4b}) can now be evolved in time by applying
the Runge-Kutta algorithm of the appropriate (third) accuracy order.

\item
A final computational step is then needed to provide the relation between
$\hb_i$ primary data and $b_i$ staggered point-value fields, those used in
flux calculations. One has then to solve the {\em implicit} relations
\be
[I-\frac{1}{24}\sD_i^{(2)}](b_i)=\hb_i,
\label{implicit}
\ee
for $i=x,y,z$, typically by means of iterative methods (see LD).
As discussed in details in Sect.~2.4, the crucial point concerning
how to measure magnetic monopoles is that {\em the same} algorithm to compute
first derivatives for fluxes should be actually applied to define the
$b_i$ first derivatives in the $\divnn$ sum of Eq.~(\ref{div2}), which is
exactly preserved {\em only} for $\hb_i$ primary data.
Thus, when derivatives of $\divnn$ are actually calculated starting
from $b_i$ fields (obtained implicitly from Eq.~(\ref{implicit})),
which is the definition actually relevant for numerical monopoles and that
is in fact measured in our tests of Sect.~4, in a similar way as shown in
Eq.~(\ref{D2}) for fluid fluxes, the $\divnn$ variable will not be zero
to within machine accuracy, though it can be made arbitrarily small
depending on the accuracy of the inversion method employed.

\end{enumerate}

In the actual implementation of the code, if one wants to keep track
of magnetic field-lines a slightly different approach,
perfectly equivalent to that outlined above, can be followed.
The induction equation (\ref{4b}) may be substituted by
\be
\frac{{\rm d}}{{\rm d}t}[A_k(t)]=E_k,
\label{4d}
\ee
where $A_k$ is the point-value representation of the vector potential
$k$ component and $E_k$ of the corresponding electric field component
(see the line-averaged counterpart, Eq.~(\ref{potential})).
The double non-oscillatory derivation in Eq.~(\ref{hE_z}) must be
now applied to, say, $A_z$ rather than to $E_z$:
\be
\hA_z=A_z-\frac{1}{24}[\Dx^{(2)}(A_z)+\Dy^{(2)}(A_z)],
\label{hA_z}
\ee
and the divergence-free staggered $\hb_i$ fields are now defined
(as in Eq.~(\ref{potrep}), proper for the finite-volume framework) as
\be
\hb_i=\sum_{j,k}\epsilon_{i,j,k} D_j(\hA_k),
\ee
to which the above remarks on $\divnn$ apply unchanged.

\clearpage

\clearpage

\begin{table}
\begin{center}
\begin{tabular}{lccccc}
\hline
   & $\ad_8$ & $\ad_{16}$ & $\ad_{32}$ & $\ad_{64}$ & $\ad_{128}$ \\
\hline
POSITIVE-UCT & 0.53097 & 0.12792 & 0.04273 & 0.01254 & 0.00322 \\
MC-HLL-UCT   & 0.60488 & 0.13133 & 0.04507 & 0.01392 & 0.00393 \\
CENO-HLL-UCT & 0.31342 & 0.03759 & 0.00461 & 0.00056 & 0.00012 \\
\hline
\end{tabular}
\end{center}
\caption{Averaged errors on the transverse velocity and magnetic field
components for the oblique 2-D CP Alfv\'en wave problem. The errors are
measured from the numerical solution at time $t=1$ (one wave period),
compared with the corresponding initial setting, for increasing
resolutions and for the various UCT schemes.}
\label{table_alfven}
\end{table}

\clearpage

\begin{table}
\begin{center}
\begin{tabular}{lcccccccc}
\hline
 & $\rho$ & $v_\xi$ & $v_\eta$ & $v_z$ & $p$ & $B_\xi$ & $B_\eta$ & $B_z$ \\
\hline
{\bf Test ST-1: L}  & 1 & 10 & 0 & 0 &  20 & $5/\sqrt{4\pi}$
 & $5/\sqrt{4\pi}$ & 0 \\
{\bf Test ST-1: R}  & 1 & -10 & 0 & 0 & 1 & $5/\sqrt{4\pi}$
 & $5/\sqrt{4\pi}$ & 0 \\
\hline
{\bf Test ST-2: L}  & 1.08 & 1.2 & 0.01 & 0.5 & 0.95 & $2/\sqrt{4\pi}$
 & $3.6/\sqrt{4\pi}$ & $2/\sqrt{4\pi}$ \\
{\bf Test ST-2: R}  & 1 & 0 & 0 & 0 & 1 & $2/\sqrt{4\pi}$
 & $4/\sqrt{4\pi}$ & $2/\sqrt{4\pi}$ \\
\hline
{\bf Test ST-3: L}  & 1 & 0 & 0 & 0 & 1 & 0.75 & 1 & 0 \\
{\bf Test ST-3: R}  & 0.125 & 0 & 0 & 0 & 0.1 & 0.75 & -1 & 0 \\
\hline
\end{tabular}
\end{center}
\caption{Constant left (L) and right (R) states for the three oblique
shock tube problems.}
\label{table_st_lr}
\end{table}


\begin{table}
\begin{center}
\begin{tabular}{lcccccc}
\hline
{\bf Test ST-1}   & $\ad$ & $\delta B_\xi$ & $[\divb]_{\rm max}$ & 
$[\divb]_{\rm avr}$ & $[\divbb]_{\rm max}$ & $[\divbb]_{\rm avr}$ \\
\hline
POSITIVE-BS & 0.0225 & 0.0051 & $0.11\times 10^3$ & $ 0.30\times 10^1$ &
- & - \\
POSITIVE-UCT & 0.0126 & 0.0019 & $0.94\times 10^2$ & $0.26\times 10^1$ & 
 $0.11\times 10^{-12}$ & $0.67\times 10^{-15}$ \\
MC-HLL-BS & 0.0304 & 0.0090 & $0.12\times 10^3$ & $0.32\times 10^1$ &
- & - \\
MC-HLL-UCT & 0.0227 & 0.0021 & $0.10\times 10^3$ & $0.25\times 10^1$ &
$0.11\times 10^{-12}$ & $0.44\times 10^{-15}$ \\
CENO-HLL-BS & 0.0320 & 0.0092 & $0.12\times 10^3$ & $0.33\times 10^1$ &
- & - \\
CENO-HLL-UCT & 0.0227 & 0.0021 & $0.11\times 10^3$ & $0.27\times 10^1$ &
$0.22\times 10^{-6}$ & $0.57\times 10^{-8}$ \\
\hline
\\
\hline
{\bf Test ST-2}   & $\ad$ & $\delta B_\xi$ & $[\divb]_{\rm max}$ & 
$[\divb]_{\rm avr}$ & $[\divbb]_{\rm max}$ & $[\divbb]_{\rm avr}$ \\
\hline
POSITIVE-BS & 0.0155 & 0.0016 & $0.65\times 10^1$ & $ 0.34\times 10^0$ &
- & - \\
POSITIVE-UCT & 0.0119 & 0.0006 & $0.70\times 10^1$ & $0.29\times 10^0$ & 
 $0.57\times 10^{-13}$ & $0.33\times 10^{-15}$ \\
MC-HLL-BS & 0.0237 & 0.0025 & $0.61\times 10^1$ & $0.35\times 10^0$ &
- & - \\
MC-HLL-UCT & 0.0200 & 0.0005 & $0.66\times 10^1$ & $0.21\times 10^0$ &
$0.57\times 10^{-13}$ & $0.22\times 10^{-15}$ \\
CENO-HLL-BS & 0.0201 & 0.0025 & $0.71\times 10^1$ & $0.41\times 10^0$ &
- & - \\
CENO-HLL-UCT & 0.0154 & 0.0006 & $0.63\times 10^1$ & $0.25\times 10^0$ &
$0.35\times 10^{-7}$ & $0.14\times 10^{-8}$ \\
\hline
\\
\hline
{\bf Test ST-3}   & $\ad$ & $\delta B_\xi$ & $[\divb]_{\rm max}$ & 
$[\divb]_{\rm avr}$ & $[\divbb]_{\rm max}$ & $[\divbb]_{\rm avr}$ \\
\hline
POSITIVE-BS & 0.0365 & 0.0043 & $0.16\times 10^2$ & $ 0.66\times 10^0$ &
- & - \\
POSITIVE-UCT & 0.0165 & 0.0005 & $0.86\times 10^1$ & $0.30\times 10^0$ & 
 $0.14\times 10^{-13}$ & $0.17\times 10^{-15}$ \\
MC-HLL-BS & 0.0586 & 0.0064 & $0.22\times 10^2$ & $0.13\times 10^1$ &
- & - \\
MC-HLL-UCT & 0.0295 & 0.0003 & $0.46\times 10^1$ & $0.19\times 10^0$ &
$0.28\times 10^{-13}$ & $0.22\times 10^{-15}$ \\
CENO-HLL-BS & 0.0612 & 0.0079 & $0.25\times 10^2$ & $0.17\times 10^1$ &
- & - \\
CENO-HLL-UCT & 0.0257 & 0.0004 & $0.65\times 10^1$ & $0.23\times 10^0$ &
$0.13\times 10^{-6}$ & $0.18\times 10^{-8}$ \\
\hline
\end{tabular}
\end{center}
\caption{Numerical errors for the three oblique shock tube problems.
For the various schemes, the errors are calculated from the 2-D $256\times 2$
run with respect to the corresponding 1-D high resolution run with $1024$
grid points. The displayed errors are the average $\ad$ L1 norm of all
the involved variables, the error on the parallel field component
$B_\xi$ (that supposed to remain constant), the maximum and averaged values
of $\divn$, and the maximum and averaged values of $\divnn$, which is not 
defined, of course, for non-CT schemes.}
\label{table_st}
\end{table}

\clearpage

\clearpage

\clearpage

\clearpage

\begin{table}
\begin{center}
\begin{tabular}{lccc}
\hline
   & $\ad_{50}$ & $\ad_{100}$ & $\ad_{200}$ \\
\hline\hline
POSITIVE-UCT & 0.1601 & 0.0780 & 0.0296 \\
MC-HLL-UCT   & 0.1898 & 0.0920 & 0.0358 \\
CENO-HLL-UCT & 0.1477 & 0.0657 & 0.0205 \\
\hline
\end{tabular}
\end{center}
\caption{Averaged L1 norms on the involved variables for the
OT vortex problem at $t=\pi$. Errors are measured against a
high resolution ($400^2$) reference run.} 
\label{table_ot}
\end{table}

\begin{table}
\begin{center}
\begin{tabular}{lccc}
\hline
   & $\ad_{50}$ & $\ad_{100}$ & $\ad_{200}$ \\
\hline
POSITIVE-UCT & 0.1751 & 0.0888 & 0.0375 \\
MC-HLL-UCT   & 0.1770 & 0.0876 & 0.0331 \\
CENO-HLL-UCT & 0.1525 & 0.0719 & 0.0239 \\
\hline
\end{tabular}
\end{center}
\caption{Averaged L1 norms on the involved variables for the
fast rotor problem at $t=0.15$. Errors are measured against a
high resolution ($400^2$) reference run.} 
\label{table_rotor}
\end{table}

\clearpage

\clearpage

\end{document}